\documentclass[journal]{IEEEtran}

\usepackage{xcolor,soul,framed}
\colorlet{shadecolor}{yellow}
\usepackage{graphicx}
\usepackage[caption=false,font=footnotesize]{subfig}
\graphicspath{{./}{./figures/main_results/quadriga_uma_nlos/}{./figures/main_results/umi_nlos/}{./figures/main_results/cdlb_qdg/}{./figures/ablation/model_variants/}{./figures/ablation/model_variants_prtn1/}{./figures/ablation/kfd/}{./figures/fewshot/}}
\DeclareGraphicsExtensions{.pdf,.jpeg,.png}

\usepackage[cmex10]{amsmath}
\usepackage{amssymb}
\usepackage{array}
\usepackage{mdwmath}
\usepackage{mdwtab}
\usepackage{eqparbox}
\usepackage{url}
\usepackage{booktabs}

\begin{document}
\title{DDA-Net: Accurate TDD Channel Estimation via Deep Unfolding the Doppler-Delay-Angle Representation of Channel Signals}
\author{Yufei Ma,
        Xu Zhu,
        and Tiejun Li%
\thanks{The authors are with LMAM and School of Mathematical Sciences, and Center for Machine Learning Research, Peking University, Beijing 100871, P.R. China. (e-mail: myf18@pku.edu.cn; xuzhu@pku.edu.cn; tieli@pku.edu.cn)}%
\thanks{Corresponding author: Tiejun Li}
}

\markboth{}{}

\makeatletter
\def\@IEEEpubidpullup{4\baselineskip}
\makeatother
\IEEEpubid{\parbox[b]{\textwidth}{\normalfont\fontsize{7}{8}\selectfont
This is the author-accepted manuscript of an article accepted for publication in \emph{IEEE Transactions on Wireless Communications}.\\[1pt]
\copyright~2026 IEEE. Personal use of this material is permitted. Permission from IEEE must be obtained for all other uses, in any current or future media, including reprinting/republishing this material for advertising or promotional purposes, creating new collective works, for resale or redistribution to servers or lists, or reuse of any copyrighted component of this work in other works.
}}

\maketitle

\begin{abstract}
In TDD massive MIMO systems, channel estimation under sparse frequency-hopping pilots is challenging: each snapshot captures only one narrow pilot block that hops across frequency, with tens of milliseconds between adjacent snapshots. Finite-window leakage and off-grid effects weaken the ideal Doppler-delay-angle (DDA) sparsity, limiting both classical sparse recovery and purely data-driven approaches lacking an explicit structured transform-domain model. We propose DDA-Net, a model-driven 3D deep unfolding network for joint multi-snapshot channel state reconstruction. DDA-Net integrates an ADMM-based formulation with a closed-form data-consistency update that avoids tensor inversion, a lightweight Doppler-domain learned prior, and delay oversampling to mitigate basis mismatch. It consistently outperforms strong baselines across three channel settings from 3GPP TR 38.901: UMa-NLOS, UMi-NLOS, and CDL-B. Ablations confirm that window-level 3D processing and explicit Doppler modeling yield target-dependent improvements. With minimal target-domain fine-tuning, the UMa-pretrained model surpasses both its zero-shot version and counterparts trained from scratch with the same number of target-domain samples. The Doppler-domain design proves consistently superior to its time-domain equivalent, with a wider margin after fine-tuning. These results demonstrate that combining exact physical data consistency with a learned DDA-domain prior is an effective and sample-efficient approach to channel state acquisition under sparse frequency-hopping pilots.
\end{abstract}


\IEEEpeerreviewmaketitle

\section{Introduction}

Time-division duplex (TDD) massive multiple-input multiple-output (MIMO) systems rely on accurate uplink channel state information (CSI) for coherent combining, spatial multiplexing, and reciprocity-based precoding~\cite{Marzetta2010MassiveMIMO,Larsson2014MassiveMIMO}. CSI acquisition becomes substantially harder when the channel evolves between pilot-bearing observations, since channel aging reduces the usefulness of stale CSI and calls for estimators that exploit temporal structure rather than treating each observation as static~\cite{Truong2013ChannelAging}. We study a channel acquisition setting in which each sounding reference signal (SRS) transmission occupies one OFDM symbol and covers one narrow contiguous pilot block; the observed block hops over frequency, and adjacent observations are separated by $\Delta t=40$~ms. For periodic and semi-persistent SRS resources, NR radio resource control (RRC) configures the periodicity and offset in slots; for the underlying 30-kHz NR numerology used to configure the SRS resources, the 80-slot option corresponds to 40~ms~\cite{GPP38331,GPP38211}. We use this NR-configurable timing together with an NR SRS bandwidth and frequency-hopping configuration that yields $K=17$ frequency positions as the main operating point. This deliberately sparse setting is motivated by capacity-oriented sounding: less frequent and narrower per-UE SRS transmissions can potentially accommodate more UEs within a fixed sounding-resource budget, at the cost of a harder full-band reconstruction problem.

\IEEEpubidadjcol

The large inter-snapshot interval narrows the unambiguous sampled-Doppler range and can lead to appreciable channel variation between adjacent observations. Consequently, the estimator must cope with weaker temporal correlation and possible sampled-Doppler aliasing rather than rely on near-static evolution across snapshots. At the same time, each individual snapshot remains severely underdetermined because only a narrow pilot block is observed, motivating joint reconstruction over a finite window.

Classical model-based studies have shown that massive-MIMO channels exhibit exploitable structure beyond a single OFDM symbol. In the space-time domain, joint estimation across successive OFDM symbols can benefit from common sparse supports~\cite{Gao2016SCSMassiveMIMO,Liu2016BurstSparsity}. In the Doppler-delay-angle (DDA) domain, OTFS-related studies have likewise shown structured sparsity and its usefulness for channel prediction~\cite{Shen2019OTFSMassiveMIMO,Yin2020PADPrediction}. These results strongly motivate a DDA representation; however, they are not enough in our setting since these methods in the space-time domain generally operate on observations separated by tens of microseconds, roughly three orders of magnitude shorter than our $40$~ms inter-snapshot interval, and therefore rely on much stronger temporal correlation than is available here.

Even with a DDA representation, sparse frequency-hopping pilots over a short window remain challenging for classical on-grid sparse models. Fractional delays and Dopplers do not align with isolated DFT atoms, so energy leaks across neighboring bins and reconstruction becomes sensitive to dictionary resolution, regularization, and window size~\cite{Chi2011BasisMismatch}. Off-grid Bayesian estimators partly alleviate this issue by explicitly modeling continuous parameters~\cite{Wei2022OffGridOTFS,Zhu2025FHSJCEP}, while gridless super-resolution via atomic norms offers a cleaner formulation at the cost of semidefinite relaxations and multilevel Toeplitz/Vandermonde machinery~\cite{Tang2013OffGrid,Yang2016MDToeplitz}. For repeated window-level CSI reconstruction, these approaches remain cumbersome in the present setting.

The closest work to our task lies in CSI acquisition and extrapolation with frequency-hopping uplink pilots. Wan and Liu studied TDD 5G NR channel extrapolation under a hopping uplink pilot pattern, Zhu~\emph{et al.} developed an off-grid message-passing method for joint estimation and prediction, and Wan~\emph{et al.} later considered multi-user pilot-pattern optimization~\cite{Wan2024TDD5GNRExtrapolation,Zhu2025FHSJCEP,Wan2025PilotPatternOptimization}. Zhu~\emph{et al.} use an off-grid DAD model to address DFT-based energy leakage and oversample the Doppler grid for future-channel prediction; in their default configuration, partial-band comb SRS transmissions visit four subbands within each full-band sounding, with adjacent full-band soundings separated by four slots. Wan and Liu instead address coherence loss across timeslots and bandwidth parts (BWPs) from time-varying imperfection factors, combining multi-timeslot, multi-BWP observations while estimating those factors and subsequently tracking continuous Doppler frequency offsets and off-grid delay/angle parameters. In contrast, each snapshot in our setting observes one contiguous $1/17$-band block, and a ten-snapshot window covers only 10 of the 17 frequency blocks; this geometry motivates delay refinement for full-band completion at the sampled sounding times rather than future-time prediction or recursive tracking. To the best of the authors' knowledge, no prior work addresses window-level full CSI reconstruction under such sparse frequency-hopping pilots and such a large inter-snapshot interval.

Learning-based methods provide a complementary direction. Early deep OFDM receivers learned implicit pilot- or symbol-to-data mappings without explicitly maintaining a structured channel representation~\cite{Ye2018PowerDL}. ChannelNet-style methods later recast channel estimation itself as image restoration on the observation grid~\cite{Soltani2018ChannelNet}, while more recent architectures such as Channelformer and LBPCE exploit attention mechanisms or 3D convolutional processing to better capture channel structure in OFDM and time-varying MIMO-OFDM settings~\cite{Luan2023Channelformer,Liu2024LBPCE}. For massive MIMO, Chun~\emph{et al.} proposed a deep estimator tailored to short-pilot regimes~\cite{Chun2019MassiveMIMODLCE}. These approaches can be powerful, but under sparse frequency-hopping pilots they must infer cross-time and cross-frequency structure from extremely incomplete observation-domain inputs: each snapshot reveals only a narrow pilot block, and the large inter-snapshot interval weakens the temporal correlation available for reliable extrapolation. Without an explicit low-dimensional transform-domain model, such methods are therefore better suited to local denoising or interpolation than to stable window-level CSI reconstruction in the present setting. Model-driven deep learning offers a middle ground between rigid iterative solvers and fully black-box regression~\cite{He2019ModelDriven,Balatsoukas2019DeepUnfolding}. Representative examples include LAMP~\cite{Borgerding2017LAMP}, LDAMP~\cite{Metzler2017LDAMP}, ADMM-CSNet~\cite{Yang2020ADMMCSNet}, and learned beamspace channel estimation via algorithm unrolling~\cite{He2018LDAMP}. However, existing unfolded channel estimators are largely 2D, mmWave/beamspace specific, or designed for denser and more regular observations than the window-level frequency-hopping setting studied here.

ADMM separates the data-fidelity, proximal, and dual updates~\cite{Boyd2011ADMM}. Deep unfolding can retain an analytic data-consistency update while representing the proximal map by a learned module~\cite{Yang2020ADMMCSNet}. Building on this framework, we tailor the state representation and operator splitting to window-level CSI reconstruction from sparse frequency-hopping pilots. All snapshots in a window are jointly represented by an $N_\theta\times N_\tau\times N_\nu$ tensor, and a 3D learned prior jointly models local angle-delay-Doppler structure and finite-window leakage in the DDA domain. At each stage, the current state is transformed to the time-delay-angle domain, where a closed-form DC update incorporates the snapshot-dependent pilot sampling operator; the updated state is then transformed back to the DDA domain. A delay-oversampled dictionary improves the representation of fractional-delay paths while preserving the row orthogonality used in the present closed-form DC derivation. These task-specific choices instantiate the generic unfolding framework as a full-window DDA parameterization with 3D prior modeling and pilot-aware domain switching.

In this study, we demonstrate that integrating a carefully chosen physical model (the DDA representation) with a data-driven deep learning strategy leads to a significant improvement in channel estimation accuracy. The main contributions of this paper are as follows.
\begin{itemize}
\item To the best of our knowledge, DDA-Net is the first deep-learning-based method for TDD massive MIMO channel estimation to use a full-window 3D DDA representation as its internal reconstruction state. We formulate the task as a 3D DDA inverse problem and develop a domain-switching ADMM-unfolded architecture: a 3D learned prior jointly processes the full-window DDA state, while an analytic DC update in the time-delay-angle domain exactly incorporates the pilot sampling operator of each snapshot.
\item We combine a lightweight residual prior with a delay-oversampled DDA representation to model local structure across angle, delay, and Doppler and to mitigate finite-window and off-grid leakage. Controlled ablations isolate the contributions of full-window 3D processing, Doppler parameterization, and delay oversampling.
\item We show that, under the same pilot overhead, the \emph{minimum-coverage-distance (MCD) pilot}, a coverage-aware design, substantially improves DDA-Net and the full-window classical solvers over the Standard hopping pilot, and that \emph{lightweight few-shot fine-tuning} enables sample-efficient adaptation under domain shift. A systematic evaluation on three 3GPP TR 38.901 channel settings, UMa-NLOS, UMi-NLOS, and CDL-B, further demonstrates that DDA-Net achieves the best overall performance across the evaluated in-distribution and target-domain settings.
\end{itemize}

The rest of this paper is organized as follows. We begin in Section~\ref{sec:SystemModel} by introducing the system model and formulating the inverse problem. Section~\ref{sec:DDA-Net} then presents our proposed method. In Section~\ref{sec:Implementation}, we detail the implementation used in our experiments. We turn to the experimental study in Section~\ref{sec:Experiments}, followed by a discussion of key design choices and limitations in Section~\ref{sec:Discussion}. Finally, we draw our conclusions.

\section{System Model and Problem Formulation}\label{sec:SystemModel}

This section introduces the system model, the observation structure under frequency-hopping pilots, and the DDA-based optimization formulation.

\subsection{Windowed Channel and Observation Model}
We consider window-level uplink CSI reconstruction from a single-antenna user equipment (UE) to a base station (BS) with an $N_{\mathrm{v}}\times N_{\mathrm{h}}$ receive array and $N_{\mathrm{pol}}$ polarization components. The total number of receive channels is therefore $N_{\mathrm{rx}}=N_{\mathrm{v}}N_{\mathrm{h}}N_{\mathrm{pol}}$.

Let $q\in\mathcal{Q}\triangleq\{0,\ldots,T_{\mathrm{w}}-1\}$ denote the discrete snapshot index within a reconstruction window. The $q$th sounding snapshot is acquired at the continuous time instant $t_q=t_0+q\Delta t$, where $t_0$ is the beginning of the window and $\Delta t$ is the inter-snapshot interval. Over $N_{\mathrm{f}}$ subcarriers, the windowed uplink channel in the time--frequency--space (TFS) domain is denoted by
\begin{equation}
\mathbf{H}\in\mathbb{C}^{N_{\mathrm{rx}}\times N_{\mathrm{f}}\times T_{\mathrm{w}}},
\quad
\mathbf{H}_q\in\mathbb{C}^{N_{\mathrm{rx}}\times N_{\mathrm{f}}},\quad q\in\mathcal{Q},
\label{eq:sec2_channel_tensor}
\end{equation}
where $\mathbf{H}_q$ is the frequency-response matrix at time $t_q$. The numerical values of $N_{\mathrm{rx}}$, $N_{\mathrm{f}}$, $T_{\mathrm{w}}$, and $\Delta t$ used in the experiments are specified in Section~\ref{sec:impl_data}.

After absorbing the known pilot symbols into the observation model, only $M_{\mathrm{p}}$ subcarriers are observed at each snapshot. We collect the resulting measurements as
\begin{equation}
\mathbf{Y}\in\mathbb{C}^{N_{\mathrm{rx}}\times M_{\mathrm{p}}\times T_{\mathrm{w}}},
\quad
\mathbf{Y}_q\in\mathbb{C}^{N_{\mathrm{rx}}\times M_{\mathrm{p}}},\quad q\in\mathcal{Q},
\label{eq:sec2_observation_tensor}
\end{equation}
and write the window-level measurement process as
\begin{equation}
\mathbf{Y}=\mathcal{M}(\mathbf{H})+\mathbf{N},
\label{eq:sec2_window_operator}
\end{equation}
where $\mathcal{M}(\cdot)$ is the frequency-hopping pilot sampling operator and $\mathbf{N}$ is additive complex Gaussian noise. Because $M_{\mathrm{p}}\ll N_{\mathrm{f}}$, each snapshot is strongly underdetermined on its own; the reconstruction therefore jointly exploits the pilot observations and shared channel structure across the window.

\subsection{Frequency-Hopping Pilot Pattern Across the Window}
The full band is partitioned into $K$ non-overlapping contiguous pilot blocks of $M_{\mathrm{p}}$ subcarriers, with $N_{\mathrm{f}}=K M_{\mathrm{p}}$. At discrete snapshot $q$, only one block is observed, and its position is determined by a configured cyclic hopping sequence. Let $\Omega_q\subset\{1,\ldots,N_{\mathrm{f}}\}$ denote the observed subcarrier set, where $|\Omega_q|=M_{\mathrm{p}}$. The per-snapshot observation model is
\begin{equation}
\mathbf{Y}_q = \mathcal{P}_{\Omega_q}(\mathbf{H}_q) + \mathbf{N}_q,
\qquad q\in\mathcal{Q},
\label{eq:sec2_physical_observation}
\end{equation}
where $\mathcal{P}_{\Omega_q}(\cdot)$ extracts the columns indexed by $\Omega_q$ and $\mathbf{N}_q\in\mathbb{C}^{N_{\mathrm{rx}}\times M_{\mathrm{p}}}$ denotes additive complex Gaussian noise with effective variance $\sigma_q^2$. Thus, the reconstruction problem combines a spectrally sparse observation at each snapshot with a sampling mask that changes across the window. The specific block partitions and hopping sequences used in the experiments are specified in Section~\ref{sec:impl_data}.

\subsection{Multipath Channel and DDA Representations}\label{sec:dda_representation}
The wideband massive MIMO channel at discrete snapshot~$q$ can be expressed as a superposition of $L$ effective propagation paths,
\begin{equation}
\mathbf{H}_q
=
\sum_{l=1}^{L}
\beta_{l,q}
\mathbf{a}_l(\boldsymbol{\theta}_{l,q})\,\mathbf{d}(\tau_{l,q})^T,
\label{eq:sec2_multipath}
\end{equation}
where $\beta_{l,q}$, $\tau_{l,q}$, and $\boldsymbol{\theta}_{l,q}=(\theta_{v,l,q},\theta_{h,l,q})$ are the complex coefficient, delay, and pair of normalized receive spatial frequencies associated with the two-dimensional angle of arrival of the $l$th path at snapshot~$q$, respectively. Snapshot-to-snapshot variations in $\tau_{l,q}$ and $\boldsymbol{\theta}_{l,q}$ represent range and angle migration, respectively, while changes in the phase increments of $\beta_{l,q}$ represent Doppler migration. For the dual-polarized $N_{\mathrm{v}}\times N_{\mathrm{h}}$ BS array,
\[
\mathbf{a}_l(\boldsymbol{\theta}_{l,q})
=
\mathbf{p}_l\otimes \mathbf{a}_{\mathrm{h}}(\theta_{h,l,q})\otimes \mathbf{a}_{\mathrm{v}}(\theta_{v,l,q}),
\]
where $\mathbf{p}_l\in\mathbb{C}^{N_{\mathrm{pol}}}$ collects the polarization weights. The component vectors satisfy
\[
\relax[\mathbf{a}_{\mathrm{v}}(\theta_v)]_m=\frac{1}{\sqrt{N_{\mathrm{v}}}}e^{-j2\pi m\theta_v},
\quad
\relax[\mathbf{a}_{\mathrm{h}}(\theta_h)]_n=\frac{1}{\sqrt{N_{\mathrm{h}}}}e^{-j2\pi n\theta_h},
\]
for $m=0,\ldots,N_{\mathrm{v}}-1$ and $n=0,\ldots,N_{\mathrm{h}}-1$. The frequency-domain delay response is $[\mathbf{d}(\tau)]_f=e^{-j2\pi f\Delta f\tau}$ for $f=0,\ldots,N_{\mathrm{f}}-1$, where $\Delta f$ is the spacing of the working frequency-sample grid; in our experiments, retaining every eighth tone of the underlying 30-kHz NR grid gives $\Delta f=240$~kHz. The dictionaries introduced below are applied at each snapshot, so the resulting angle-delay states retain the snapshot-dependent channel evolution.

To obtain the DDA representation, we introduce a receiver-side space-angle dictionary
\begin{equation}
\mathbf{F}_{\mathrm{sa}}
\;=\;
\mathbf{I}_{N_{\mathrm{pol}}}\otimes\left(\mathbf{F}_{\mathrm{h}}\otimes \mathbf{F}_{\mathrm{v}}\right)
\in\mathbb{C}^{N_{\mathrm{rx}}\times N_\theta},
\label{eq:sec2_spatial_angle_dictionary}
\end{equation}
where $\mathbf{F}_{\mathrm{v}}\in\mathbb{C}^{N_{\mathrm{v}}\times (N_{\mathrm{v}} k_{\mathrm{sa}})}$ and $\mathbf{F}_{\mathrm{h}}\in\mathbb{C}^{N_{\mathrm{h}}\times (N_{\mathrm{h}} k_{\mathrm{sa}})}$ are DFT steering dictionaries obtained by uniformly sampling $\mathbf{a}_{\mathrm{v}}(\cdot)$ and $\mathbf{a}_{\mathrm{h}}(\cdot)$, so $N_\theta=N_{\mathrm{pol}}N_{\mathrm{v}}N_{\mathrm{h}} k_{\mathrm{sa}}^2$. With $k_{\mathrm{sa}}=1$, we have $N_\theta=N_{\mathrm{rx}}$, so $\mathbf{F}_{\mathrm{sa}}$ is square and unitary. This unitarity is important later because it enables a closed-form data-consistency step.

Along the frequency axis, we use an oversampled delay dictionary
\begin{equation}
\mathbf{F}_{\mathrm{fd}}=
\frac{1}{\sqrt{N_\tau}}\mathbf{S}_{\mathrm{f}}\mathbf{W}_{N_\tau}^{H}
\in \mathbb{C}^{N_{\mathrm{f}}\times N_\tau},
\quad
N_\tau = k_{\mathrm{fd}} N_{\mathrm{f}},
\label{eq:sec2_delay_dictionary}
\end{equation}
where $\mathbf{W}_{N_\tau}$ is the unnormalized $N_\tau$-point DFT matrix, $\mathbf{S}_{\mathrm{f}}=[\mathbf{I}_{N_{\mathrm{f}}}\ \mathbf{0}] \in \{0,1\}^{N_{\mathrm{f}}\times N_\tau}$ selects the first $N_{\mathrm{f}}$ rows, and $k_{\mathrm{fd}}$ is the delay oversampling factor. Equivalently, the rows of $\mathbf{F}_{\mathrm{fd}}^{H}$ are sampled delay atoms $\mathbf{d}(\bar{\tau}_n)^T/\sqrt{N_\tau}$ on the grid $\bar{\tau}_n=n/(N_\tau\Delta f)$. The corresponding time-delay-angle representation is denoted by
\begin{equation}
\mathbf{X}\in\mathbb{C}^{N_\theta\times N_\tau\times T_{\mathrm{w}}},
\quad
\mathbf{X}_q\in\mathbb{C}^{N_\theta\times N_\tau},\quad q\in\mathcal{Q},
\label{eq:sec2_angle_delay_time}
\end{equation}
and is related to the original channel through the synthesis model
\begin{equation}
\mathbf{H}_q = \mathbf{F}_{\mathrm{sa}}\mathbf{X}_q\mathbf{F}_{\mathrm{fd}}^H.
\label{eq:sec2_synthesis_relation}
\end{equation}
Because $\mathbf{F}_{\mathrm{sa}}$ is unitary and $\mathbf{F}_{\mathrm{fd}}$ has orthonormal rows, we may define $\mathbf{X}_q=\mathbf{F}_{\mathrm{sa}}^H\mathbf{H}_q\mathbf{F}_{\mathrm{fd}}$. With this definition, $\mathbf{H}_q$ is recovered exactly through~\eqref{eq:sec2_synthesis_relation}. The modeling approximation is therefore the expected localization of the transform-domain state rather than the transform itself: continuous off-grid multipath broadens the angle-delay support through basis mismatch~\cite{Shen2019OTFSMassiveMIMO,Chi2011BasisMismatch}. Oversampling in delay ($k_{\mathrm{fd}}>1$) mitigates the delay mismatch while preserving the row-orthonormality of the per-snapshot sensing matrix later defined in~\eqref{eq:sec2_sensing_matrix}, namely~\eqref{eq:sec2_sensing_row_orth}, which is needed for the closed-form DC update.

\begin{figure*}[!t]
\centering
\subfloat[Space--frequency and angle--delay.]{
    \includegraphics[width=0.310\textwidth]{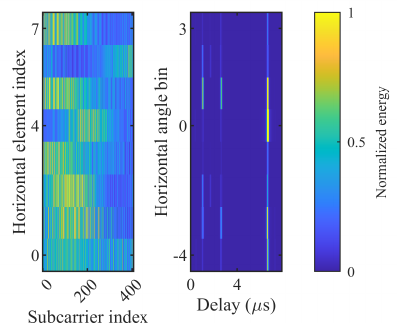}
}
\hfill
\subfloat[Delay--time.]{
    \includegraphics[width=0.310\textwidth]{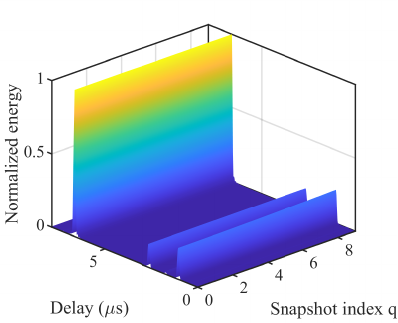}
}
\hfill
\subfloat[Delay--Doppler.]{
    \includegraphics[width=0.310\textwidth]{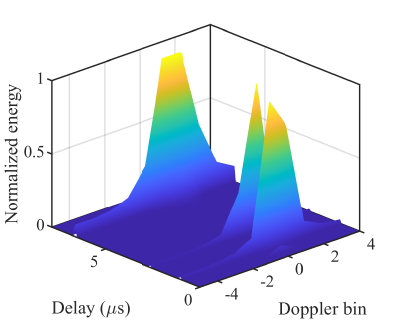}
}
\caption{Transform-domain structure of one held-out full-band UMa-NLOS window with $T_{\mathrm{w}}=N_\nu=10$ and $\Delta t=40$~ms; no pilot-based reconstruction or learned estimator is involved. Panel (a) shows the center snapshot in the space--frequency (left) and angle--delay (right) domains; each view is normalized by its $99.5$th-percentile energy and saturated at one on the shared linear color scale. All delay-domain views use the default $k_{\mathrm{fd}}=3$ oversampling to refine the delay grid. Panels (b)--(c) show the angle-marginal energy before and after the unitary $T_{\mathrm{w}}$-point DFT along $q$; except for the third-axis coordinate, they use identical delay ranges, viewpoints, and display scaling based on a joint $99.5$th-percentile reference, so the comparison isolates the temporal DFT. For this window, retaining $90\%$ of the third-axis marginal energy requires 9 of 10 snapshot indices in (b), but only 4 of 10 Doppler bins in (c). These counts are computed from the original unclipped energy; near-zero displayed values indicate weak support.}
\label{fig:dda_structure}
\end{figure*}

In the DDA model, we further represent temporal structure in the Doppler domain by applying a unitary $T_{\mathrm{w}}$-point DFT along the discrete snapshot axis. Since the present formulation uses the full-window transform without Doppler truncation, we have $N_\nu=T_{\mathrm{w}}$. The corresponding DDA representation is
\begin{equation}
\widetilde{\mathbf{X}}
 = \mathrm{FFT}_{q\rightarrow \nu}(\mathbf{X})
 \in \mathbb{C}^{N_\theta\times N_\tau\times N_\nu},
\label{eq:sec2_doppler_representation}
\end{equation}
with inverse relation $\mathbf{X}=\mathrm{IFFT}_{\nu\rightarrow q}(\widetilde{\mathbf{X}})$. Fig.~\ref{fig:dda_structure} visualizes this representation chain using full-band ground truth. These transforms do not add observations; instead, they reparameterize the channel so that most of its energy is carried by localized angle-delay neighborhoods, while the temporal DFT concentrates the third-axis marginal into fewer Doppler bins. Under the incomplete observation model in~\eqref{eq:sec2_physical_observation}, this compressibility reduces the effective number of significant degrees of freedom presented to the learned prior and aligns with the local processing of its 3D convolutional module, which can regularize clustered support and finite leakage rather than dense snapshot-wise variation. The subsequent comparisons and controlled ablations further provide empirical evidence for the benefit of exploiting this transform-domain structure. Recent ST-DDA analysis further shows that exact within-window constancy of the complex path coefficient and Doppler shift is unnecessary for a useful DDA representation: when the path-coefficient variation rate and phase curvature are sufficiently small, their variation induces only limited additional Doppler spreading within a finite window~\cite{Zhu2026STDDA}. Accordingly, DDA-Net's 3D learned prior relies on finite-window compressibility rather than exact path stationarity: the common grid records evolving paths as local trajectories across the time-delay-angle snapshots and as a concentrated support region with finite spreading after the temporal DFT, while more pronounced range, angle, or Doppler migration can broaden the support and make prior modeling more difficult.

To clarify the sampled-Doppler ambiguity, consider the locally constant-parameter special case
\[
\beta_{l,q}=\beta_l e^{j2\pi\nu_lq\Delta t},\qquad
\tau_{l,q}=\tau_l,\qquad
\boldsymbol{\theta}_{l,q}=\boldsymbol{\theta}_l,
\]
where $\beta_l$ is the reference complex coefficient and $\nu_l$ is the physical Doppler shift in hertz. On the uniform sounding grid $t_q=t_0+q\Delta t$, for any integer $m$,
\[
e^{j2\pi(\nu_l+m/\Delta t)q\Delta t}
=
e^{j2\pi\nu_lq\Delta t},
\]
so continuous physical Doppler is identifiable only modulo $1/\Delta t$. Our target, however, is the full channel $\{\mathbf{H}_q\}$ at the sampled sounding times rather than a unique continuous path Doppler. Under this special case, alias-equivalent Dopplers generate the same sampled channel and therefore correspond to the same reconstruction target. Recovering the unobserved subcarriers still relies on complementary pilot observations and learned channel structure, but it does not require selecting a unique member of the continuous-Doppler alias class. Accordingly, the present task is sampled-time full-band channel completion rather than continuous-Doppler de-aliasing.

\subsection{Optimization Problem}
Define the snapshot-dependent sensing matrix
\begin{equation}
\mathbf{A}_q = [\mathbf{F}_{\mathrm{fd}}]_{\Omega_q,:}
\in \mathbb{C}^{M_{\mathrm{p}}\times N_\tau},
\label{eq:sec2_sensing_matrix}
\end{equation}
that is, the row submatrix of $\mathbf{F}_{\mathrm{fd}}$ indexed by the observed pilot set $\Omega_q$. Then~\eqref{eq:sec2_physical_observation} and~\eqref{eq:sec2_synthesis_relation} yield the equivalent virtual-domain observation model
\begin{equation}
\mathbf{Y}_q = \mathbf{F}_{\mathrm{sa}}\mathbf{X}_q\mathbf{A}_q^H + \mathbf{N}_q.
\label{eq:sec2_virtual_observation}
\end{equation}
Because $\mathbf{F}_{\mathrm{sa}}$ is unitary in the present $k_{\mathrm{sa}}=1$ setting and each $\mathbf{A}_q$ is formed by selecting rows from $\mathbf{F}_{\mathrm{fd}}$, the rows of $\mathbf{A}_q$ remain orthonormal:
\begin{equation}
\mathbf{A}_q\mathbf{A}_q^H=\mathbf{I}_{M_{\mathrm{p}}}.
\label{eq:sec2_sensing_row_orth}
\end{equation}
Together, the unitarity of $\mathbf{F}_{\mathrm{sa}}$ and the row orthonormality of $\mathbf{A}_q$ enable the closed-form DC update used in DDA-Net.

Given this observation model, a natural objective for window-level reconstruction is
\begin{equation}
\begin{aligned}
\min_{\mathbf{X}} \quad
& \sum_{q\in\mathcal{Q}}
\frac{1}{2\sigma_q^2}
\left\|
\mathbf{Y}_q-\mathbf{F}_{\mathrm{sa}}\mathbf{X}_q\mathbf{A}_q^H
\right\|_F^2
\;+\;
\lambda\,\Phi\!\left(\widetilde{\mathbf{X}}\right) \\
\text{with}\quad
& \widetilde{\mathbf{X}}=\mathrm{FFT}_{q\rightarrow \nu}(\mathbf{X}),\ \ \mathbf{X}=\mathrm{IFFT}_{\nu\rightarrow q}(\widetilde{\mathbf{X}})
\end{aligned}
\label{eq:sec2_regularized_problem}
\end{equation}
where the first term measures data fidelity under the physical sensing model, and $\Phi(\cdot)$ denotes a transform-domain prior that promotes structure in the DDA representation. Writing the prior on $\widetilde{\mathbf{X}}$ rather than directly on $\mathbf{X}$ is motivated by the fact that dynamic multipath is often more localized after the temporal DFT, although the same inverse problem can also be parameterized in the time-delay-angle domain. We therefore reparameterize~\eqref{eq:sec2_regularized_problem} in terms of $\widetilde{\mathbf{X}}$ and introduce an auxiliary variable $\widetilde{\mathbf{Z}}$ to obtain the ADMM splitting
\begin{equation}
\text{(DDA Model)}\quad\quad\quad
\begin{aligned}
\min_{\widetilde{\mathbf{X}},\widetilde{\mathbf{Z}}} \quad
& f_{\mathrm{DC}}(\widetilde{\mathbf{X}})
\;+\;
g(\widetilde{\mathbf{Z}}) \\
\text{s.t.}\quad
& \widetilde{\mathbf{X}}=\widetilde{\mathbf{Z}},
\end{aligned}
\label{eq:sec2_admm_split}
\end{equation}
where $f_{\mathrm{DC}}(\widetilde{\mathbf{X}})$ denotes the data-fidelity term in~\eqref{eq:sec2_regularized_problem}, evaluated after mapping $\widetilde{\mathbf{X}}$ back to the time domain by an inverse temporal DFT, and $g(\widetilde{\mathbf{Z}})=\lambda\Phi(\widetilde{\mathbf{Z}})$ denotes the prior term. This splitting separates data consistency under the physical sensing model from transform-domain prior regularization. Let $\widetilde{\boldsymbol{\beta}}$ denote the scaled dual variable. For a penalty parameter $\rho>0$, the conventional scaled-form ADMM iterations for~\eqref{eq:sec2_admm_split} are
\begin{subequations}\label{eq:sec2_conventional_admm}
\begin{align}
\widetilde{\mathbf{X}}^{(n)}
&=
\mathop{\arg\min}_{\widetilde{\mathbf{X}}}
f_{\mathrm{DC}}(\widetilde{\mathbf{X}})
+\frac{\rho}{2}
\left\|
\widetilde{\mathbf{X}}
-\widetilde{\mathbf{Z}}^{(n-1)}
+\widetilde{\boldsymbol{\beta}}^{(n-1)}
\right\|_F^2,
\label{eq:sec2_conventional_admm_x}
\\
\widetilde{\mathbf{Z}}^{(n)}
&=
\operatorname{prox}_{g/\rho}
\left(
\widetilde{\mathbf{X}}^{(n)}
+\widetilde{\boldsymbol{\beta}}^{(n-1)}
\right),
\label{eq:sec2_conventional_admm_z}
\\
\widetilde{\boldsymbol{\beta}}^{(n)}
&=
\widetilde{\boldsymbol{\beta}}^{(n-1)}
+\widetilde{\mathbf{X}}^{(n)}
-\widetilde{\mathbf{Z}}^{(n)}.
\label{eq:sec2_conventional_admm_beta}
\end{align}
\end{subequations}
These three steps are, respectively, the data-consistency update, the proximal update associated with $g$, and the scaled-dual update driven by the equality-constraint residual. They provide the direct template for the unfolded DDA-Net stages described in Section~\ref{sec:DDA-Net}.

\section{DDA-Net Architecture}\label{sec:DDA-Net}

This section presents the DDA-Net architecture, detailing the unfolding of ADMM iterations into a feed-forward network with closed-form data consistency, a learned prior module, and delay-domain oversampling.

\subsection{From 3D ADMM to DDA-Net}
DDA-Net truncates and unfolds the three ADMM updates in~\eqref{eq:sec2_conventional_admm} into a finite-depth network. At stage $n$, the $\widetilde{\mathbf{X}}$ update remains an analytic DC step defined by the snapshot-specific pilot operators: $\widetilde{\mathbf{Z}}^{(n-1)}-\widetilde{\boldsymbol{\beta}}^{(n-1)}$ is transformed by IFFT to the time-delay-angle domain, updated in closed form for each snapshot, and transformed by FFT back to the DDA domain. The proximal update is represented by a stage-specific learned surrogate $\mathcal{D}_{\phi_n}$, while $\gamma$ controls the dual-update step size. Accordingly, $\widetilde{\mathbf{X}}$, $\widetilde{\mathbf{Z}}$, and $\widetilde{\boldsymbol{\beta}}$ retain their roles as the main, auxiliary, and scaled-dual states, respectively, all represented as full-window 3D DDA tensors.

The fixed elements are the pilot index sets $\Omega_q$, the dictionaries $\mathbf{F}_{\mathrm{sa}}$ and $\mathbf{F}_{\mathrm{fd}}$, the sensing matrices $\mathbf{A}_q$, the FFT/IFFT operations, and the analytic form of the DC update. The penalty $\rho$ and dual step size $\gamma$ are trainable scalars shared across stages. The stage-specific parameters $\phi_n$ of $\mathcal{D}_{\phi_n}$ comprise Conv3D weights and biases and B-spline coefficients, and implicitly encode the prior $g=\lambda\Phi$ and its effective regularization strength. The three ADMM variables serve as data-dependent stage states. The architecture comprises one initialization stage, $N_{\mathrm{iter}}$ intermediate stages, and one terminal unfolded stage, followed by a DC-only output layer. Thus, the main configuration with $N_{\mathrm{iter}}=8$ contains $L=N_{\mathrm{iter}}+2=10$ stages with learned-prior and dual updates.

\begin{figure*}[t]
\centering
\includegraphics[width=\textwidth]{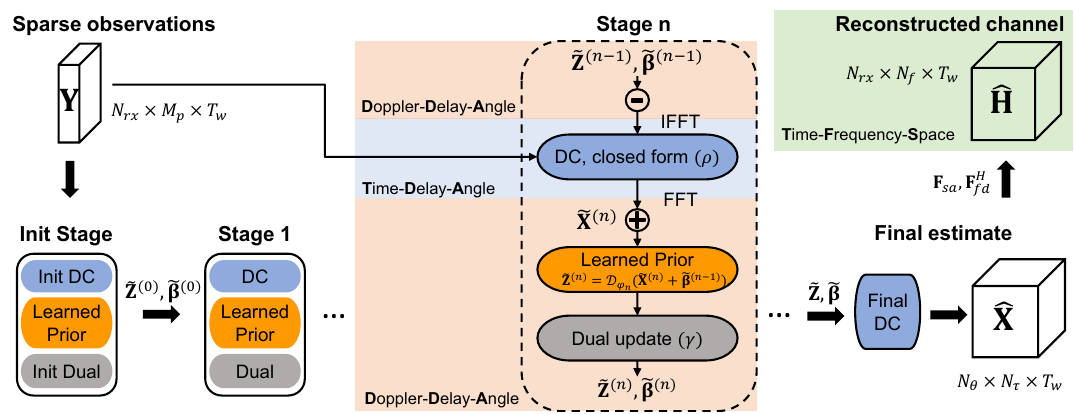}
\caption{Overview of DDA-Net and one representative unfolded stage. An initialization stage is followed by $N_{\mathrm{iter}}$ intermediate stages, one terminal unfolded stage, and a DC-only output layer. Within each unfolded stage, the closed-form DC update operates in the time-delay-angle domain with penalty $\rho$, the learned prior $\mathcal{D}_{\phi_n}$ operates in the DDA domain, and the scaled-dual update uses step size $\gamma$; IFFT/FFT switches between the two domains. The trainable scalars $\rho$ and $\gamma$ are shared across stages, whereas $\phi_n$ is stage specific; the sensing operators and domain transforms are fixed, and $\widetilde{\mathbf{X}}$, $\widetilde{\mathbf{Z}}$, and $\widetilde{\boldsymbol{\beta}}$ are full-window, data-dependent DDA states. The final time-delay-angle output $\hat{\mathbf{X}}$ is mapped to the TFS domain to obtain $\hat{\mathbf{H}}$.}
\label{fig:ddanet_architecture}
\end{figure*}

\subsection{Closed-Form Data Consistency Update}
The central model-based component of DDA-Net is the DC update. In the Doppler-domain version, the center of the quadratic DC penalty from the previous iteration is first mapped to the time-delay-angle domain:
\begin{equation}
\mathbf{V}^{(n)}_{\mathrm{time}}
=
\mathrm{IFFT}_{\nu\rightarrow q}
\left(
\widetilde{\mathbf{Z}}^{(n-1)}-\widetilde{\boldsymbol{\beta}}^{(n-1)}
\right).
\label{eq:sec3_v_time}
\end{equation}
For each snapshot index $q\in\mathcal{Q}$, let $\mathbf{V}^{(n)}_q\in\mathbb{C}^{N_\theta\times N_\tau}$ denote the corresponding slice. With $\rho>0$ denoting the ADMM penalty parameter, the DC subproblem reduces to
\begin{equation}
\begin{aligned}
\mathbf{X}_q^{(n)}
=\mathop{\arg\min}_{\mathbf{X}_q}\quad
&\frac{1}{2\sigma_q^2}
\left\|
\mathbf{Y}_q-\mathbf{F}_{\mathrm{sa}}\mathbf{X}_q\mathbf{A}_q^H
\right\|_F^2 \\
&+\frac{\rho}{2}
\left\|
\mathbf{X}_q-\mathbf{V}^{(n)}_q
\right\|_F^2.
\end{aligned}
\label{eq:sec3_dc_subproblem}
\end{equation}
Because $\mathbf{F}_{\mathrm{sa}}^H\mathbf{F}_{\mathrm{sa}}=\mathbf{I}$ in the current setting and $\mathbf{A}_q\mathbf{A}_q^H=\mathbf{I}_{M_{\mathrm{p}}}$, this problem admits the closed-form solution
\begin{equation}
\mathbf{X}_q^{(n)}
=
\mathbf{C}_q^{(n)}
-
\frac{1}{\rho\sigma_q^2+1}
\mathbf{C}_q^{(n)}\mathbf{A}_q^H\mathbf{A}_q,
\label{eq:sec3_dc_closed_form}
\end{equation}
with
\begin{equation}
\mathbf{C}_q^{(n)}
=
\mathbf{V}_q^{(n)}
+
\frac{\mathbf{F}_{\mathrm{sa}}^H\mathbf{Y}_q\mathbf{A}_q}{\rho\sigma_q^2}.
\label{eq:sec3_dc_c}
\end{equation}
Equation~\eqref{eq:sec3_dc_closed_form} enforces model-based data consistency in DDA-Net by explicitly incorporating the snapshot-specific pilot operator, rather than requiring the network to learn the measurement process implicitly. The initial DC layer is obtained by setting this quadratic-penalty center to zero, while the final DC layer uses the most recent auxiliary and dual states to produce the time-delay-angle output. The derivation of~\eqref{eq:sec3_dc_closed_form}--\eqref{eq:sec3_dc_c}, including the simplification of the inverse term, is given in Appendix~\ref{app:dc_derivation}.

\subsection{DDA-Domain Iteration}
Based on the ADMM splitting~\eqref{eq:sec2_admm_split}, DDA-Net maintains its persistent stage states in the DDA domain while carrying out the closed-form DC step in the time-delay-angle domain. One unfolded iteration can be written as
\begin{align}
\mathbf{V}^{(n)}_{\mathrm{time}}
&=
\mathrm{IFFT}_{\nu\rightarrow q}
\left(
\widetilde{\mathbf{Z}}^{(n-1)}-\widetilde{\boldsymbol{\beta}}^{(n-1)}
\right),
\label{eq:sec3_doppler_step1}
\\
\mathbf{X}^{(n)}_{\mathrm{time}}
&=
\mathrm{DC}\!\left(
\mathbf{V}^{(n)}_{\mathrm{time}},\mathbf{Y}
\right),
\label{eq:sec3_doppler_step2}
\\
\text{(DDA-Net)}\quad \widetilde{\mathbf{X}}^{(n)}
&=
\mathrm{FFT}_{q\rightarrow \nu}
\left(
\mathbf{X}^{(n)}_{\mathrm{time}}
\right),
\label{eq:sec3_doppler_step3}
\\
\widetilde{\mathbf{Z}}^{(n)}
&=
\mathcal{D}_{\phi_n}
\left(
\widetilde{\mathbf{X}}^{(n)}+\widetilde{\boldsymbol{\beta}}^{(n-1)}
\right),
\label{eq:sec3_doppler_step4}
\\
\widetilde{\boldsymbol{\beta}}^{(n)}
&=
\widetilde{\boldsymbol{\beta}}^{(n-1)}
+
\gamma
\left(
\widetilde{\mathbf{X}}^{(n)}-\widetilde{\mathbf{Z}}^{(n)}
\right).
\label{eq:sec3_doppler_step5}
\end{align}
The Doppler-domain parameterization makes temporal structure explicit by transforming the snapshot axis into Doppler. Over the finite observation window, this representation provides a structured axis for modeling slowly evolving multipath, while the snapshot-dependent sampling masks supply complementary measurements across the window. An alternative \emph{time-domain} parameterization keeps all internal variables in $\mathbb{C}^{N_\theta\times N_\tau\times T_{\mathrm{w}}}$ and applies the prior directly to the time-domain tensor $\mathbf{X}$, removing the FFT/IFFT conversion steps in~\eqref{eq:sec3_doppler_step1} and~\eqref{eq:sec3_doppler_step3}. This variant shares the same DC update and dictionaries but must learn temporal coupling implicitly through local 3D convolutions, making it close in spirit to an ADMM-CSNet style unfolding~\cite{Yang2020ADMMCSNet}. It is retained as a controlled ablation in Section~\ref{sec:ablation_model}.

\subsection{Learned Prior Module}
The learned prior module serves as the surrogate for the $\mathbf{Z}$-proximal update in each unfolded stage. Under the Doppler-domain parameterization, it maps $\mathbf{U}=\widetilde{\mathbf{X}}^{(n)}+\widetilde{\boldsymbol{\beta}}^{(n-1)}$ to $\widetilde{\mathbf{Z}}^{(n)}$; under the time-domain parameterization, it maps $\mathbf{U}=\mathbf{X}^{(n)}+\boldsymbol{\beta}^{(n-1)}$ to $\mathbf{Z}^{(n)}$. In the Doppler-domain version, $\widetilde{\mathbf{Z}}^{(n)}-\widetilde{\boldsymbol{\beta}}^{(n)}$ centers the next DC update, while $\widetilde{\mathbf{X}}^{(n)}-\widetilde{\mathbf{Z}}^{(n)}$ is the residual driving the dual update. The analytic DC branch explicitly incorporates the measurement model through the data-fidelity term, whereas the learned-prior branch models structure in the active internal domain.
\begin{equation}
\mathcal{D}_{\phi_n}(\mathbf{U})
=
\mathbf{U}
-
\mathcal{C}_{2,\phi_n}
\Big(
\Psi_{\phi_n}
\big(
\mathcal{C}_{1,\phi_n}(\mathbf{U})
\big)
\Big),
\label{eq:sec3_denoiser}
\end{equation}
where $\mathcal{C}_{1,\phi_n}$ and $\mathcal{C}_{2,\phi_n}$ are 3D convolutions and $\Psi_{\phi_n}$ is a channel-wise B-spline nonlinearity. The first Conv3D extracts local features in the active internal domain, the channel-wise B-spline applies a trainable nonlinearity, and the second Conv3D maps these features to a residual correction that is subtracted from $\mathbf{U}$. This compact prior follows the lightweight denoising block used in ADMM-CSNet~\cite{Yang2020ADMMCSNet} and is embedded here in a 3D, window-level, domain-switching unfolded architecture.

Complex tensors are represented by separate real and imaginary channels for the real-valued 3D convolutions, and the B-spline acts separately on the real and imaginary parts of each feature channel. Padding along the third axis differs between the two parameterizations: the Doppler-domain version uses circular padding along the Doppler axis to reflect DFT periodicity, whereas the time-domain version uses zero padding along the snapshot axis. Before the B-spline, each sample and feature channel is scaled so that its maximum magnitude over the three tensor dimensions matches the B-spline range; the scale is restored after the activation. The remaining implementation details are given in Section~\ref{sec:Implementation}.

\subsection{Delay-Domain Oversampling and Training Loss}

Delay-domain oversampling sets the dictionary size to $N_\tau=k_{\mathrm{fd}}N_{\mathrm{f}}$ and thereby refines the virtual representation without altering the physical observation model. We prioritize the delay dimension because of the present observation geometry. At each sounding instant, all $N_{\mathrm{rx}}$ receive channels are observed, but only on one contiguous $M_{\mathrm{p}}$-tone block. In the main 40-ms, $K=17$ configuration, $M_{\mathrm{p}}/N_{\mathrm{f}}=24/408=1/17$ and $T_{\mathrm{w}}=10<K=17$, so a window cannot cover all frequency blocks, whereas the reconstruction target is the full-band channel at those same sampled sounding times. The missing responses therefore lie directly along the frequency axis. Because continuous path delays determine the phase evolution across subcarriers, fractional-delay mismatch directly degrades recovery of these unobserved frequency responses, motivating delay-grid refinement in the present configuration.

Delay oversampling also preserves the closed-form structure of the unfolded DC step. For any $k_{\mathrm{fd}}$, the sensing matrix $\mathbf{A}_q$ formed from rows of the oversampled DFT dictionary satisfies $\mathbf{A}_q\mathbf{A}_q^H=\mathbf{I}$, so the same closed-form DC update remains valid while the internal delay dimension $N_\tau$ grows linearly with $k_{\mathrm{fd}}$. Additional refinement of the angle or Doppler grids would further enlarge the 3D state and requires separate accuracy--cost tradeoffs. We therefore retain the native angle grid and finite-window Doppler transform, and evaluate $k_{\mathrm{fd}}\in\{1,2,3\}$ in Section~\ref{sec:Experiments} to quantify the accuracy--cost tradeoff of delay refinement; $k_{\mathrm{fd}}=3$ is used as the default high-accuracy setting.

The network is trained against the reconstruction error in the original TFS domain. After the final DC output $\hat{\mathbf{X}}$, we reconstruct
\begin{equation}
\hat{\mathbf{H}}_q
=
\mathbf{F}_{\mathrm{sa}}\hat{\mathbf{X}}_q\mathbf{F}_{\mathrm{fd}}^H,
\qquad
q\in\mathcal{Q},
\label{eq:sec3_reconstruction_to_original}
\end{equation}
and optimize the NMSE-aligned objective
\begin{equation}
\mathcal{L}
=
\mathbb{E}
\left[
\frac{\|\hat{\mathbf{H}}-\mathbf{H}\|_F^2}{\|\mathbf{H}\|_F^2}
\right].
\label{eq:sec3_nmse_loss}
\end{equation}
This loss directly supervises the physically meaningful end task of reconstructing the full channel in the original TFS domain, rather than an intermediate transform-domain representation.

\section{Implementation Details}\label{sec:Implementation}

This section describes the data generation, network configuration, training protocol, and baseline methods used in the experiments.

\subsection{Data, Pilot, Window, and Scenario Configuration}\label{sec:impl_data}
\begin{figure}[t]
\centering
\includegraphics[width=\linewidth]{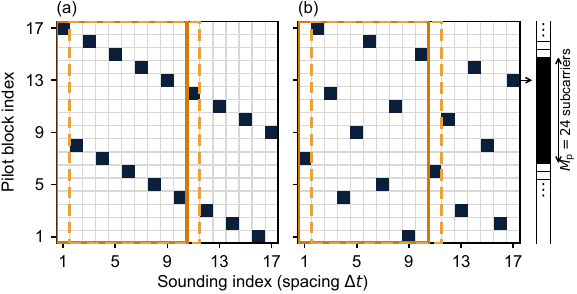}
\caption{Pilot-pattern schematics used in the experiments: (a) Standard hopping pilot; (b) MCD pilot. Each filled square represents a contiguous block of $M_{\mathrm{p}}=24$ observed subcarriers at one sounding instant, and adjacent soundings are separated by $\Delta t$. The solid and dashed rectangles mark two length-$T_{\mathrm{w}}=10$ reconstruction windows shifted by one sounding.}
\label{fig:pilot_pattern_schematic}
\end{figure}

All three channel datasets are generated using QuaDRiGa~\cite{Jaeckel2014QuaDRiGa} and follow 3GPP TR 38.901~\cite{GPP38901}. The in-distribution dataset uses the scenario-based UMa-NLOS model. We consider a single-antenna UE and a BS with a $4\times 8$ dual-polarized array, i.e., $N_{\mathrm{rx}}=64$, at carrier frequency $3.5$~GHz. The BS position is fixed at height $20$~m, while the UE initial position is randomized on a circle of radius $100$~m, the UE height is drawn from $1.2$--$1.5$~m, and the user speed is drawn from $1$--$4$~km/h. The original full-band response is sampled on $3264$ physical tones with 30-kHz NR subcarrier spacing over $97.92$~MHz. To reduce dimensionality, we retain every eighth tone, yielding a working grid of $N_{\mathrm{f}}=408$ frequency samples with effective spacing $\Delta f=240$~kHz for all reconstruction experiments. Each sample in the main setting consists of $T_{\mathrm{w}}=10$ consecutive snapshots separated by $\Delta t=40$~ms on this grid. We use $T_{\mathrm{w}}=10$ as a practical balance: jointly processing multiple hopping instants provides complementary observations for reconstruction, whereas a longer window spans more pronounced channel variation and incurs higher computational cost. The dataset contains $10,000$ base channel samples before pilot-offset expansion, split into training, validation, and test subsets (see Section~\ref{sec:impl_training}). The main pilot configuration partitions the bandwidth into $K=17$ contiguous blocks of size $M_{\mathrm{p}}=24$ and reveals one block per snapshot. Unless otherwise stated, the subsequent experiments use this 40-ms, $K=17$ main setting. The Standard hopping pilot in Fig.~\ref{fig:pilot_pattern_schematic}(a) follows the 17-state cyclic order generated by the NR SRS frequency-hopping rule specified in 3GPP TS~38.211~\cite{GPP38211}. As shown in~\cite{Zhu2026MCCPilot}, the MCD pilot, which spreads the pilot block positions as evenly as possible over the time-frequency grid, can significantly improve channel recovery accuracy under the same per-snapshot pilot budget. We therefore evaluate the MCD pilot in the present deep-learning framework using the pattern shown in Fig.~\ref{fig:pilot_pattern_schematic}(b).

To distinguish the effects of the sounding interval and frequency observation configuration, we compare four UMa-NLOS settings at $10$~dB SNR: $(\Delta t,K,M_{\mathrm{p}})=(5~\mathrm{ms},4,102)$ and $(40~\mathrm{ms},4,102)$ under the Standard hopping pilot, and $(40~\mathrm{ms},17,24)$ under both Standard and MCD pilots. For the underlying 30-kHz NR numerology, the 5- and 40-ms intervals correspond to the NR 10- and 80-slot SRS periodicity options, respectively~\cite{GPP38331,GPP38211}. All four settings retain the same $N_{\mathrm{f}}=408$ working grid and $T_{\mathrm{w}}=10$ snapshots. The two $K=4$ settings compare sounding intervals at the same block size and hopping rule, while the two $K=17$ settings compare pilot placement at the same interval and observation budget. The remaining experiments retain the main 40-ms, $K=17$ configuration.

Unless otherwise stated, the delay dictionary is oversampled by the factor $k_{\mathrm{fd}}=3$. For both out-of-distribution targets, the carrier frequency, bandwidth, antenna array, UE placement and mobility distributions, 40-ms snapshot spacing, 10-snapshot window, and pilot configurations are matched to the UMa-NLOS setting. The UMi-NLOS target retains the scenario-based 3GPP construction but switches from the urban-macro NLOS scenario to the urban-micro street-canyon NLOS scenario with a 10-m BS height, whereas the CDL-B target uses the standardized NR CDL-B profile~\cite{GPP38901}. For CDL-B, the RMS delay spread is drawn uniformly from 200 to 700~ns.

\subsection{Network Configuration}\label{sec:impl_network}
The main DDA-Net configuration uses $L=10$ unfolded stages with learned-prior and dual updates: one initialization stage, $N_{\mathrm{iter}}=8$ intermediate stages, and one terminal stage, followed by a DC-only output layer. Each learned-prior module is a lightweight residual mapping of the form Conv3D--B-spline--Conv3D with real/imaginary channel splitting and amplitude normalization. The ten learned-prior modules use independent parameters $\{\phi_n\}_{n=0}^{9}$, whereas the trainable DC penalty $\rho$ and dual step size $\gamma$ are shared across the unfolded stages. The default hidden width is $16$ channels after the real/imaginary split, and the default 3D kernel is $3\times 11\times 3$, where the larger kernel along the delay axis reflects the higher internal dimensionality in that direction. A complexity comparison with the baselines is provided in Table~\ref{tab:complexity}.

\subsection{Training Protocol}\label{sec:impl_training}
The dataset is split into training, validation, and test subsets with ratio $0.7/0.15/0.15$ using a fixed random seed. Unless otherwise noted, the full-data learning-based models are trained under mixed SNR over $\{-10,-5,0,5,\ldots,30\}$~dB. For the UMa-NLOS sounding-regime comparison, each learning-based method is trained separately for each regime using the corresponding channel sequences and pilot configuration. Because different sliding windows expose different local pilot-block configurations within the hopping cycle, training mixes all $K$ cyclic starting offsets rather than fixing a single observation mask. At test time, unless otherwise stated, all methods within each dataset and sounding setting are evaluated on the same held-out set of $500$ base samples. For each sample, NMSE is averaged in the linear domain over all $K$ possible pilot starting offsets and then over the test samples before conversion to dB. The tunable classical baselines are optimized separately for each setting on held-out development samples. All full-data learning-based models are trained on a single NVIDIA Tesla V100 GPU using Adam and early stopping based on the validation NMSE. Full hyperparameter details are provided in Appendix~\ref{app:training_details}. Few-shot fine-tuning on UMi-NLOS and CDL-B initializes each model variant from its corresponding UMa-NLOS checkpoint, uses $20$ target-domain training samples, and retains the pretrained architecture.

\subsection{Baselines and Evaluation Metric}\label{sec:impl_baselines}
We compare DDA-Net with least-squares (LS), the fast iterative shrinkage-thresholding algorithm (FISTA)~\cite{Beck2009FISTA}, ADMM~\cite{Boyd2011ADMM}, Off-Grid-MS HMP~\cite{Zhu2025FHSJCEP}, ChannelNet~\cite{Soltani2018ChannelNet}, and LDAMP~\cite{Metzler2017LDAMP}. LS serves as a transparent per-snapshot linear reference. FISTA and ADMM operate on the full window with an $\ell_1$ prior in the Doppler domain; their hyperparameters (regularization weight, step size, and iteration count) are individually swept on a small held-out development subset to minimize NMSE. Our estimation-only adaptation of Off-Grid-MS HMP omits the prediction stage and retains off-grid learning in delay, the two spatial-frequency coordinates, and Doppler under the present observation operator. Its numerical settings are tuned on held-out development samples, and its iteration count is fixed for each evaluation setting. ChannelNet is extended to a 3D CNN that processes the full TFS window, whereas LDAMP retains its native 2D slice-wise recursion with a DnCNN denoiser and is applied independently to each time slice. Extending its Onsager-corrected recursion and denoiser to a full 3D window-level model would require a substantial redesign and a markedly higher computational budget, making the comparison less controlled. Throughout the paper, the primary metric is NMSE in dB computed on the reconstructed full window in the original TFS domain.

\section{Numerical Experiments}\label{sec:Experiments}

This section evaluates DDA-Net in terms of in-distribution accuracy, out-of-distribution generalization across deployment scenarios and channel models, ablation studies, computational cost, and few-shot adaptation.

\subsection{Main Results on UMa-NLOS}
\begin{figure*}[t]
\centering
\begin{minipage}[t]{0.49\textwidth}
    \centering
    \includegraphics[width=\linewidth]{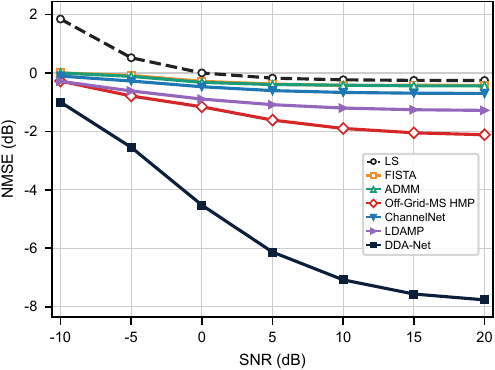}
    \\[-1mm]
    {\footnotesize (a) Standard hopping pilot.}
\end{minipage}\hfill
\begin{minipage}[t]{0.49\textwidth}
    \centering
    \includegraphics[width=\linewidth]{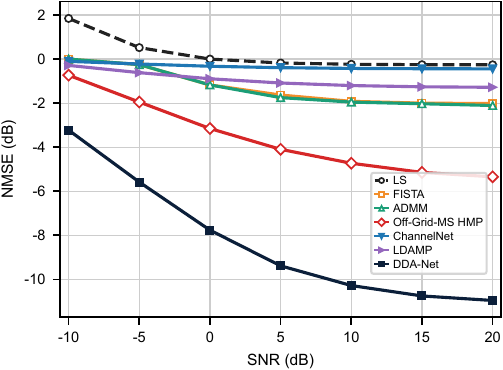}
    \\[-1mm]
    {\footnotesize (b) MCD pilot.}
\end{minipage}
\caption{Main NMSE results on the UMa-NLOS test set under (a) the Standard hopping pilot and (b) the MCD pilot. The paired panels illustrate the effect of pilot placement on estimator accuracy over the evaluated SNR range. Lower is better.}
\label{fig:uma_main_results}
\end{figure*}

Fig.~\ref{fig:uma_main_results} summarizes the in-distribution UMa-NLOS results under both pilot configurations (Section~\ref{sec:impl_data}) over test SNRs from $-10$ to $20$~dB. Testing under both pilot patterns allows us to verify that the observed gains are not specific to a single pilot geometry, and in particular to assess whether a more coverage-aware pilot arrangement improves reconstruction under the same observation budget. In both cases, DDA-Net yields the lowest NMSE across the plotted SNR range, and the gap widens as SNR increases. This trend is consistent with the nature of the task: once observation noise becomes moderate, the dominant limitation is no longer denoising but the ability to exploit the structured yet highly incomplete measurements induced by sparse frequency-hopping pilots. The proposed window-level 3D unfolding is more effective in this setting than either classical sparse recovery or observation-domain learning. In particular, the substantially weaker results of the 3D ChannelNet baseline suggest that observation-domain learning alone does not adequately exploit the virtual-domain structure underlying the extremely sparse TFS observations in this setting.

Under the Standard hopping pilot in Fig.~\ref{fig:uma_main_results}(a), DDA-Net achieves a clear advantage from $0$~dB onward. At $10$~dB SNR, DDA-Net achieves an NMSE of $-7.08$~dB, which is $5.18$~dB lower than that of HMP ($-1.90$~dB), the strongest competing baseline. FISTA, ADMM, and ChannelNet remain above $-0.90$~dB at this point, suggesting that the sparse hopping observations leave substantial ambiguity for both hand-crafted recovery methods and direct observation-domain learning. LDAMP outperforms FISTA and ADMM at medium-to-high SNRs, consistent with the benefit of its learned denoiser; however, its native 2D slice-wise formulation cannot exploit cross-time coupling within the window.

The MCD pilot in Fig.~\ref{fig:uma_main_results}(b) yields lower NMSE for the full-window classical solvers at medium-to-high SNRs. At $10$~dB SNR, HMP is the best competing method at $-4.73$~dB, while DDA-Net reaches $-10.28$~dB and retains a $5.55$~dB margin; FISTA and ADMM reach $-1.92$ and $-1.95$~dB, respectively. These results show that DDA-Net's advantage is not limited to a single pilot design and that the MCD pilot, whose smaller coverage distance provides more balanced frequency coverage within each reconstruction window, improves the conditioning of the inverse problem for methods that exploit the full observation window.

\paragraph{Comparison of sounding interval, block configuration, and pilot pattern}
Table~\ref{tab:sounding_regimes} compares these three factors at $10$~dB SNR. With $K=4$, increasing $\Delta t$ from $5$ to $40$~ms changes HMP from $-11.96$ to $-11.78$~dB, and the FISTA and ADMM results remain close. DDA-Net changes from $-18.24$ to $-15.55$~dB and remains the best method in both settings, while ChannelNet changes from $-14.28$ to $-1.48$~dB. Thus, sensitivity to the sounding interval depends on the estimator. ChannelNet operates directly in the TFS domain without an explicit DDA representation, which can make it more sensitive to the reduced temporal correlation at $40$~ms.

\begin{table}[t]
\centering
\caption{UMa-NLOS NMSE (dB) at $10$~dB SNR. All columns use $T_{\mathrm{w}}=10$, $500$ test channels, and all pilot offsets. Bold/underline: best/best competing.}
\label{tab:sounding_regimes}
\footnotesize
\setlength{\tabcolsep}{2.2pt}
\begin{tabular}{lrrrr}
\toprule
Method & \shortstack{5~ms\\$K=4$\\Standard} & \shortstack{40~ms\\$K=4$\\Standard} & \shortstack{40~ms\\$K=17$\\Standard} & \shortstack{40~ms\\$K=17$\\MCD} \\
\midrule
LS & $-1.11$ & $-1.11$ & $-0.24$ & $-0.24$ \\
FISTA & $-5.71$ & $-5.69$ & $-0.43$ & $-1.92$ \\
ADMM & $-5.94$ & $-5.79$ & $-0.42$ & $-1.95$ \\
Off-Grid-MS HMP & $-11.96$ & $\underline{-11.78}$ & $\underline{-1.90}$ & $\underline{-4.73}$ \\
ChannelNet & $\underline{-14.28}$ & $-1.48$ & $-0.67$ & $-0.42$ \\
LDAMP & $-3.07$ & $-3.26$ & $-1.20$ & $-1.20$ \\
\textbf{DDA-Net} & $\mathbf{-18.24}$ & $\mathbf{-15.55}$ & $\mathbf{-7.08}$ & $\mathbf{-10.28}$ \\
\bottomrule
\end{tabular}
\end{table}

At the same $40$-ms interval, changing from $K=4$ to $K=17$ under the Standard pilot reduces the per-snapshot sounding block from $102$ to $24$ frequency samples. Both windows still contain ten observations, but the $K=4$ window covers and revisits all four blocks, whereas the $K=17$ window visits only ten of the seventeen blocks without completing a hopping cycle. The latter therefore supplies fewer frequency measurements and less complementary or repeated coverage within the joint estimation window. HMP degrades from $-11.78$ to $-1.90$~dB, while DDA-Net changes from $-15.55$ to $-7.08$~dB. This comparison reflects the combined effects of narrower sounding bandwidth and limited-window frequency coverage.

Finally, keeping $K=17$, $\Delta t=40$~ms, $T_{\mathrm{w}}=10$, and the per-snapshot observation budget fixed, the MCD pilot improves HMP from $-1.90$ to $-4.73$~dB and DDA-Net from $-7.08$ to $-10.28$~dB. This highlights the role of the observation pattern beyond its sample count, consistent with the coverage-aware pilot-design principle in~\cite{Zhu2026MCCPilot}: nearby time--frequency pilot support is especially important when only a short sub-window is used. Off-grid refinement reduces basis mismatch, while the pilot geometry determines how strongly the available measurements constrain the candidate channel components.

\subsection{Out-of-Distribution Generalization across Deployment Scenarios and Channel Models}
\label{sec:ood_generalization}
UMi-NLOS retains the scenario-based 3GPP construction used for UMa-NLOS but changes the deployment from an urban macrocell to an urban-micro street canyon, thereby testing cross-deployment transfer~\cite{GPP38901}. CDL-B instead uses a standardized NLOS delay-angle cluster profile, thereby testing cross-model transfer~\cite{GPP38901}. The two targets therefore probe complementary OOD shifts in deployment-conditioned statistics and channel-model construction.

For both target settings, all learning-based methods trained on UMa-NLOS are evaluated directly without retraining. For each target, the classical baselines are tuned on a separate development subset and then frozen; all methods are subsequently evaluated on the same held-out test set.

\subsubsection{UMi-NLOS}

\begin{figure*}[t]
\centering
\begin{minipage}[t]{0.49\textwidth}
    \centering
    \includegraphics[width=\linewidth]{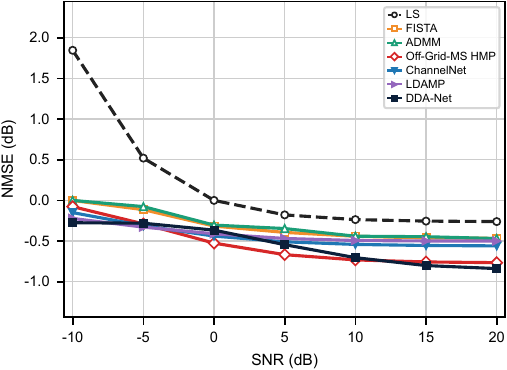}
    \\[-1mm]
    {\footnotesize (a) Standard hopping pilot.}
\end{minipage}\hfill
\begin{minipage}[t]{0.49\textwidth}
    \centering
    \includegraphics[width=\linewidth]{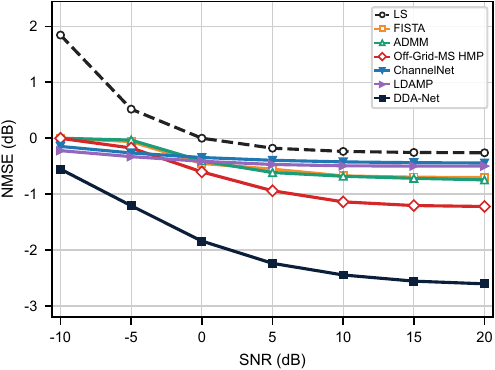}
    \\[-1mm]
    {\footnotesize (b) MCD pilot.}
\end{minipage}
\caption{Out-of-distribution NMSE results on the UMi-NLOS test set under (a) the Standard hopping pilot and (b) the MCD pilot. All learning-based methods are trained on UMa-NLOS and evaluated without retraining. The paired panels evaluate cross-deployment transfer and the effect of coverage-aware pilot placement without target-domain training. Lower is better.}
\label{fig:umi_main_results}
\end{figure*}

Fig.~\ref{fig:umi_main_results} shows the UMi-NLOS results under the two pilot settings. Under the Standard hopping pilot, DDA-Net and HMP have nearly identical average NMSE over the seven evaluated SNR points, with SNR-dependent rankings. At $10$~dB SNR, their NMSEs are $-0.70$ and $-0.73$~dB, respectively. Under the MCD pilot, DDA-Net achieves the lowest NMSE at all seven SNR points. At $10$~dB SNR, DDA-Net reaches $-2.45$~dB, compared with $-1.14$~dB for HMP, yielding a margin of $1.31$~dB. At the same SNR, switching from Standard hopping to MCD improves DDA-Net by about $1.74$~dB, compared with about $0.41$~dB for HMP.

\subsubsection{CDL-B}

\begin{figure*}[t]
\centering
\begin{minipage}[t]{0.49\textwidth}
    \centering
    \includegraphics[width=\linewidth]{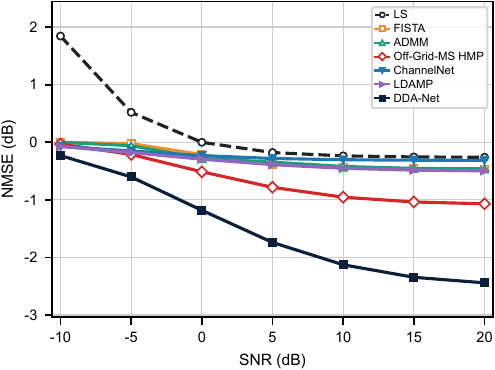}
    \\[-1mm]
    {\footnotesize (a) Standard hopping pilot.}
\end{minipage}\hfill
\begin{minipage}[t]{0.49\textwidth}
    \centering
    \includegraphics[width=\linewidth]{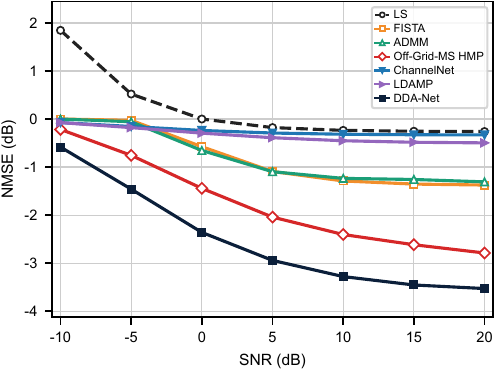}
    \\[-1mm]
    {\footnotesize (b) MCD pilot.}
\end{minipage}
\caption{Out-of-distribution NMSE results on the CDL-B test set under (a) the Standard hopping pilot and (b) the MCD pilot. All learning-based methods are trained on UMa-NLOS and evaluated without retraining. The paired panels evaluate transfer from scenario-based UMa-NLOS channels to the standardized CDL-B cluster profile and the effect of coverage-aware pilot placement under channel-model shift. Lower is better.}
\label{fig:cdlb_main_results}
\end{figure*}

Fig.~\ref{fig:cdlb_main_results} shows the corresponding results on CDL-B. DDA-Net achieves the lowest NMSE at every evaluated SNR point under both pilot settings. At $10$~dB SNR, it reaches $-2.13$~dB under the Standard hopping pilot and $-3.28$~dB under the MCD pilot. HMP is the strongest competing method at this SNR, reaching $-0.95$ and $-2.40$~dB, which leaves DDA-Net margins of $1.17$ and $0.88$~dB, respectively. Among the remaining baselines, the full-window FISTA and ADMM solvers also benefit from the MCD pilot, improving by about $0.23$/$0.24$~dB on UMi-NLOS and $0.86$/$0.82$~dB on CDL-B at $10$~dB SNR, whereas the slice-wise LDAMP baseline remains essentially unchanged.

For DDA-Net, the NMSE on UMi-NLOS at $10$~dB SNR is approximately $1.42$ and $0.83$~dB higher than that on CDL-B under the Standard and MCD pilots, respectively. Taken together, the two evaluations show that DDA-Net provides the best overall performance across the target domains without target-domain training, with a clearer advantage under the MCD pilot. These results motivate the few-shot study in Section~\ref{sec:fewshot}, which examines whether a small target-domain training set can further adapt the UMa-pretrained model.

\subsection{Ablation: Temporal Parameterization and Window-Level Processing}
\label{sec:ablation_model}
This ablation compares three model variants under the MCD pilot. \texttt{Doppler-3D} is the proposed model. \texttt{Time-3D} replaces the Doppler-domain internal representation with its time-domain counterpart while keeping the 3D denoiser and all other components unchanged, thereby isolating the effect of temporal parameterization. \texttt{Time-2D} further removes cross-snapshot coupling by processing each snapshot with a 2D denoiser, closer to the per-slice treatment in ADMM-CSNet~\cite{Yang2020ADMMCSNet}, thereby isolating the effect of window-level cross-snapshot processing.

\begin{figure*}[t]
\centering
\includegraphics[width=\textwidth]{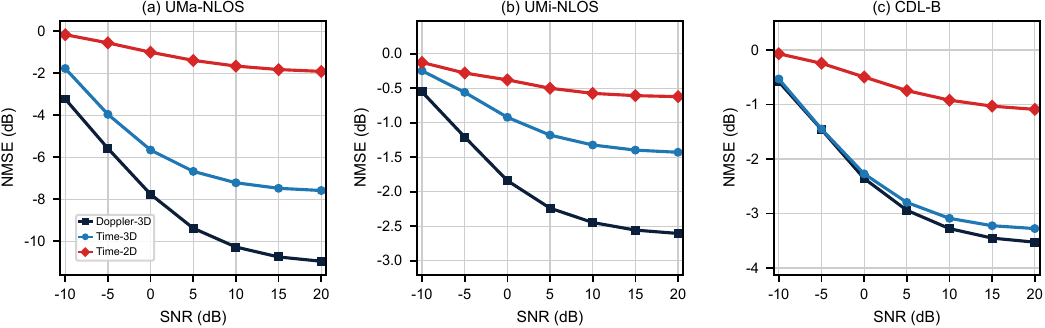}
\caption{Ablation study of temporal parameterization and window-level processing under the MCD pilot. Panels (a)--(c) show results for UMa-NLOS, UMi-NLOS, and CDL-B, respectively. \texttt{Doppler-3D} is the proposed model, \texttt{Time-3D} retains 3D cross-snapshot processing in the time domain, and \texttt{Time-2D} processes each snapshot independently. The comparisons isolate the gain from window-level cross-snapshot processing and the additional gain from explicit Doppler-domain parameterization. Lower is better.}
\label{fig:ablation_model_variants}
\end{figure*}

Fig.~\ref{fig:ablation_model_variants} shows that on UMa-NLOS at $10$~dB SNR, \texttt{Doppler-3D}, \texttt{Time-3D}, and \texttt{Time-2D} reach $-10.28$, $-7.22$, and $-1.66$~dB, respectively. Because each snapshot contains only one narrow pilot block, \texttt{Time-2D} cannot access complementary cross-snapshot observations for full-band reconstruction. Aggregating these observations with \texttt{Time-3D} improves NMSE by $5.56$~dB, and explicit Doppler-domain parameterization provides another $3.06$~dB, showing that DDA-domain organization adds value beyond 3D processing alone.

On UMi-NLOS at the same SNR, the three variants reach $-2.45$, $-1.32$, and $-0.57$~dB, respectively. The corresponding gains are $0.75$~dB from window-level processing and a further $1.12$~dB from Doppler parameterization; their different magnitudes from UMa-NLOS show that target-domain statistics affect how effectively the same observation geometry is exploited.

On CDL-B, \texttt{Time-3D} improves over \texttt{Time-2D} by $2.17$~dB, whereas \texttt{Doppler-3D} improves over \texttt{Time-3D} by only $0.19$~dB ($-3.28$ versus $-3.09$~dB). These target-dependent gaps reveal distinct transfer behavior. UMi-NLOS retains the source domain's scenario-based 3GPP construction and primarily changes deployment-conditioned statistics, which is consistent with stronger transfer of the Doppler-domain prior. CDL-B instead introduces a cross-model shift through its standardized delay-angle cluster profile; its large 3D gain but small Doppler gain indicates that cross-snapshot dependence remains exploitable, while the source-learned Doppler-domain prior is less well aligned. As discussed in Section~\ref{sec:fewshot}, the FT20 results in Fig.~\ref{fig:fewshot} further show that the weak zero-shot Doppler advantage on CDL-B is partly recoverable through target-domain calibration rather than reflecting an absence of useful Doppler structure.

\subsection{Ablation: Delay Oversampling Factor}
This ablation varies the delay oversampling factor $k_{\mathrm{fd}}\in\{1,2,3\}$ under the MCD pilot. $k_{\mathrm{fd}}=1$ corresponds to the original delay grid with no oversampling, while larger values refine the delay dictionary at the cost of increased internal dimensionality.

\begin{figure*}[t]
\centering
\includegraphics[width=\textwidth]{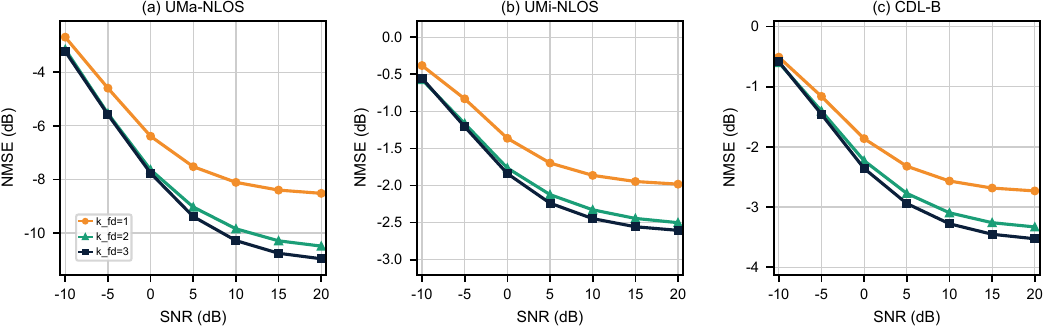}
\caption{Ablation study of the delay oversampling factor under the MCD pilot. Panels (a)--(c) show results for UMa-NLOS, UMi-NLOS, and CDL-B, respectively. The curves quantify the accuracy gained by refining the delay grid and reveal diminishing returns as $k_{\mathrm{fd}}$ increases; $k_{\mathrm{fd}}=3$ is the default configuration. Lower is better.}
\label{fig:ablation_kfd}
\end{figure*}

Fig.~\ref{fig:ablation_kfd} shows diminishing returns on all three domains. At $10$~dB SNR, the $k_{\mathrm{fd}}=1,2,3$ results are $-8.11$, $-9.84$, and $-10.28$~dB on UMa-NLOS; $-1.86$, $-2.33$, and $-2.45$~dB on UMi-NLOS; and $-2.57$, $-3.10$, and $-3.28$~dB on CDL-B. The total gains are approximately $2.17$, $0.58$, and $0.71$~dB, respectively. Because all domains use the same observation configuration and delay dictionary, the smaller OOD gains show that delay-grid refinement removes only a limited part of the target-domain error. Their similar magnitudes suggest that delay-grid mismatch is shared by UMi-NLOS and CDL-B, rather than causing their different transfer behavior. For $k_{\mathrm{fd}}=1,2,3$, batch-size-one FP32 reconstruction requires $19.5$, $39.1$, and $58.6$~G MACs, respectively. The measured peak allocated GPU memory for the network forward pass is $203$, $372$, and $542$~MiB, respectively; the parameter count remains $68.7$~K. We therefore retain $k_{\mathrm{fd}}=3$ as the default high-accuracy setting.

\subsection{Computational Cost}
Table~\ref{tab:complexity} reports model size and computational cost per ten-snapshot reconstruction window at $10$~dB SNR on UMa-NLOS, using the same configurations that produce the corresponding accuracy results in Fig.~\ref{fig:uma_main_results}. For the Standard/MCD pilots, the selected iteration counts are $40/200$ for FISTA, $80/200$ for ADMM, and $240/320$ for HMP, respectively. The predefined hyperparameter sweep uses $20$ held-out base samples and all pilot starting offsets, with NMSE as the primary selection metric. Costs are counted from the pilot observations to the reconstructed full-band channel, with complex arithmetic, FFTs, and elementwise operations expressed using a consistent approximate real-MAC-equivalent convention. Fixed sensing dictionaries and their spectral norms are precomputed. DDA-Net uses roughly one-tenth as many parameters as either learning-based baseline. Its fixed $58.6$~G MAC cost is lower than those of HMP, ChannelNet, and LDAMP under both pilot patterns, but higher than those of LS, FISTA, and ADMM. The low parameter count follows from the lightweight Conv3D--B-spline--Conv3D denoiser design with only $16$ hidden channels.

\begin{table}[t]
\centering
\caption{Model size and MACs per ten-snapshot reconstruction window on UMa-NLOS at $10$~dB SNR.}
\label{tab:complexity}
\begin{tabular}{lrrr}
\toprule
Method & Params & \multicolumn{2}{c}{MACs (G)} \\
\cmidrule(lr){3-4}
 & & Standard & MCD \\
\midrule
LS & -- & 0.0 & 0.0 \\
FISTA & -- & 8.5 & 36.6 \\
ADMM & -- & 16.2 & 37.7 \\
Off-Grid-MS HMP & -- & 327.4 & 247.8 \\
\midrule
ChannelNet 3D & 725 K & 189.4 & 189.4 \\
LDAMP (10 unrolls) & 664 K & 348.4 & 348.4 \\
\textbf{DDA-Net (ours)} & 68.7 K & 58.6 & 58.6 \\
\bottomrule
\end{tabular}
\end{table}

\subsection{Few-Shot Target-Domain Adaptation}\label{sec:fewshot}
\begin{figure*}[t]
\centering
\includegraphics[width=\textwidth]{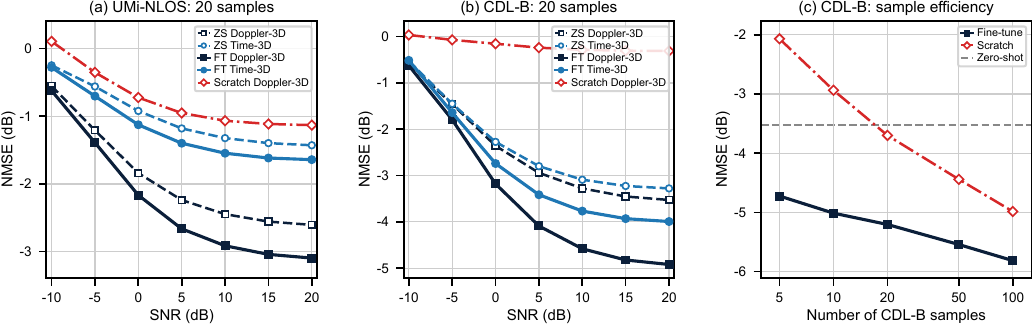}
\caption{Few-shot target-domain adaptation under the MCD pilot. Panels (a) and (b) compare Doppler-3D and Time-3D under zero-shot transfer and $20$-sample fine-tuning on UMi-NLOS and CDL-B, together with a $20$-sample Doppler-3D scratch control. Panel (c) compares fine-tuning and training from scratch on CDL-B for target-domain budgets of $5$--$100$ samples, with training and evaluation fixed at $20$~dB SNR; the dashed line denotes the zero-shot Doppler-3D baseline. Lower is better.}
\label{fig:fewshot}
\end{figure*}

We next study few-shot target-domain adaptation under the MCD pilot. For both UMi-NLOS and CDL-B, Doppler-3D and Time-3D are initialized from their corresponding UMa-NLOS pretrained checkpoints and fine-tuned with $20$ target-domain training samples. Each fine-tuned model is compared with its zero-shot counterpart; for Doppler-3D, we also include the same architecture trained from scratch using the same $20$ samples. The $20$-sample fine-tuned and scratch models are evaluated on the same target-domain test sets of $500$ base samples as their zero-shot counterparts. Fig.~\ref{fig:fewshot}(a) and (b) report these comparisons, while panel (c) varies the CDL-B training budget from $5$ to $100$ samples, with training and evaluation fixed at $20$~dB SNR.

For Doppler-3D on UMi-NLOS, fine-tuning improves upon the zero-shot model at all seven evaluated SNR points, with an average gain of $0.35$~dB. It also outperforms the model trained from scratch at all seven points by an average of $1.52$~dB. At $10$~dB SNR, the NMSE values of the fine-tuned, zero-shot, and scratch models are $-2.91$, $-2.45$, and $-1.07$~dB, respectively. On CDL-B, the corresponding average gains are $0.92$~dB over zero-shot and $3.24$~dB over training from scratch; at $10$~dB SNR, the three models reach $-4.58$, $-3.28$, and $-0.29$~dB, respectively. These within-domain comparisons show that UMa-NLOS pretraining provides an effective initialization for both cross-deployment and cross-model adaptation: at the same $20$-sample budget, fine-tuning consistently outperforms training from scratch.

Cross-target evaluation of the $20$-sample fine-tuned models (UMi-FT20 and CDL-B-FT20) reveals asymmetric adaptation. Averaged over the seven SNRs, UMi-FT20 is $0.13$~dB worse than the shared UMa-NLOS pretrained checkpoint when evaluated on CDL-B, whereas CDL-B-FT20 is $0.70$~dB worse when evaluated on UMi-NLOS. Together with the larger in-domain gain on CDL-B, this pattern suggests that CDL-B fine-tuning produces a more target-specific model, whereas UMi-NLOS fine-tuning preserves more cross-target performance.

Fig.~\ref{fig:fewshot}(a) and (b) also compare the two temporal parameterizations. On UMi-NLOS, the seven-SNR average advantage of Doppler-3D over Time-3D is $0.91$~dB under zero-shot transfer and $1.08$~dB after FT20. On CDL-B, the corresponding averages are $0.14$ and $0.57$~dB. Thus, the Doppler-domain representation provides a transferable structural bias across both target domains, and limited target-domain calibration strengthens its advantage in each case.

Fig.~\ref{fig:fewshot}(c) further quantifies sample efficiency on CDL-B at $20$~dB SNR. Fine-tuning with only $5$ samples reaches $-4.72$~dB, improving upon the zero-shot model by $1.20$~dB and the corresponding scratch model by $2.65$~dB. With $10$ samples, fine-tuning reaches $-5.01$~dB, essentially matching the $-4.99$~dB result of the scratch model trained with $100$ samples. At the $100$-sample budget, fine-tuning reaches $-5.81$~dB, maintaining a $0.83$~dB advantage over training from scratch. On CDL-B, the consistent advantage at matched sample budgets shows that source-domain pretraining substantially reduces the amount of target-domain data required for adaptation.

\section{Discussion}\label{sec:Discussion}

This section summarizes the key design insights, discusses the role of delay oversampling, and outlines the limitations and future directions of this work.

\subsection{Design Insights}
DDA-Net combines analytic data consistency with learned DDA-domain prior modeling. Under sparse frequency-hopping pilots, window-level Time-3D processing aggregates frequency blocks observed at different snapshots, improving over Time-2D by $5.56$, $0.75$, and $2.17$~dB on UMa-NLOS, UMi-NLOS, and CDL-B, respectively. With the observation protocol matched across domains, these unequal gains indicate that the benefit of aggregating the same cross-snapshot measurements depends on target-domain channel statistics. The sounding-regime and fixed-speed results further characterize reconstruction behavior across different observation densities and channel-evolution regimes.

The corresponding gains from explicit Doppler-domain parameterization are $3.06$, $1.12$, and $0.19$~dB before adaptation. The larger UMi-NLOS gain is consistent with its retaining the source domain's scenario-based 3GPP construction, whereas CDL-B introduces a cross-model shift through its standardized cluster profile. Because the physical observation model is unchanged, the wider Doppler-versus-time gap after FT20 indicates that target-domain calibration improves the learned Doppler-domain prior. Accordingly, the analytic data-consistency update retains its form, while the learned prior and other trainable unfolding parameters are initialized from the source domain and calibrated to the target domain.

\subsection{Delay Oversampling}
Delay oversampling improves all three domains, but the smaller and similar OOD gains suggest that it addresses a shared fractional-delay limitation rather than their different temporal-transfer behavior. Because refinement proportionally increases memory and computation, we retain $k_{\mathrm{fd}}=3$ as the default high-accuracy setting; adaptive or gridless representations remain future work.

\paragraph{Off-grid refinement under sparse hopping}
Off-grid learning in HMP reduces basis mismatch by refining the delay, spatial-frequency, and Doppler coordinates. With off-grid learning enabled, HMP exhibits limited reconstruction accuracy for $K=17$, particularly under the Standard pilot with $T_{\mathrm{w}}=10$. Each snapshot then contains $24$ frequency samples, and the joint window covers ten of the seventeen hopping blocks. Regular block hopping can also induce strong correlations between delay--Doppler atoms in the observation dictionary, as analyzed in~\cite{Zhu2026MCCPilot}. Different delay--Doppler components can therefore produce similar responses at the observed pilot positions, making joint support identification and coefficient estimation more difficult. This interpretation is consistent with Table~\ref{tab:sounding_regimes}: the wider blocks and repeated coverage for $K=4$ support accurate HMP estimation even at $40$~ms, while the more balanced MCD pattern improves its accuracy for $K=17$.

\subsection{Limitations and Future Directions}
The results should be interpreted within two main scope boundaries. First, the evaluation uses one array, bandwidth, and window configuration across three simulation-based domains: UMa-NLOS, UMi-NLOS, and CDL-B. The cross-domain and few-shot results support transfer within these settings, while broader system configurations, deployment scenarios, and channel models remain to be evaluated. Second, DDA-Net reconstructs the full channel at the sampled sounding times rather than continuous path Dopplers. Alias-equivalent Dopplers remain physically indistinguishable on the sampling grid; Appendix~\ref{app:speed_sensitivity} nevertheless shows robust reconstruction across the evaluated unaliased low-mobility range. Generalization to substantially different mobility and $\nu\Delta t$ distributions requires further study. Beyond these scope boundaries, exploiting overlap between successive windows through warm starts and extending the current single-user formulation to joint multi-user scheduling and reconstruction are natural directions for future work.

\section{Conclusion}
This paper addressed window-level uplink CSI reconstruction from sparse frequency-hopping pilots with large inter-snapshot intervals. In the considered operating regime, each snapshot observes only one narrow frequency block, making full-band reconstruction severely ill-posed. We proposed DDA-Net, a model-driven 3D deep unfolding network that combines exact closed-form data consistency in the time-delay-angle domain with learned prior modeling in the DDA domain, thereby unifying physical interpretability and data-driven correction within a single unfolding framework. To improve robustness under short-window leakage and off-grid mismatch, we further combined a learned residual prior with delay-domain oversampling while preserving the algebraic structure required for exact closed-form DC updates. Experiments show that coverage-aware hopping improves DDA-Net and the full-window classical solvers at the same pilot overhead, while few-shot target-domain calibration improves transfer across both deployment and channel-model shifts. Across the evaluated SNRs, DDA-Net provides the strongest overall results on UMa-NLOS and the two zero-shot OOD targets, UMi-NLOS and CDL-B. For Doppler-3D, limited target-domain fine-tuning consistently improves upon both zero-shot transfer and matched-budget training from scratch; after fine-tuning, its advantage over Time-3D widens on both targets. These results support model-driven DDA reconstruction as an effective and sample-efficient framework for channel acquisition under sparse frequency-hopping pilots.

\section*{Acknowledgment}
The authors acknowledge the support from National Key R\&D Program of China under grant 2021YFA1003301, and National Science Foundation of China under grant 12288101. They also thank the High-performance Computing Platform of Peking University for providing computational resources.

\appendices
\renewcommand{\theequation}{\thesection.\arabic{equation}}

\section{Closed-Form Data Consistency Derivation}\label{app:dc_derivation}
\setcounter{equation}{0}
This appendix derives the closed-form DC update in~\eqref{eq:sec3_dc_closed_form}--\eqref{eq:sec3_dc_c}. For each snapshot index $q$, the DC subproblem~\eqref{eq:sec3_dc_subproblem} is
\begin{equation}
\min_{\mathbf{X}_q}\ \frac{1}{2\sigma_q^2}\left\|\mathbf{Y}_q-\mathbf{F}_{\mathrm{sa}}\mathbf{X}_q\mathbf{A}_q^H\right\|_F^2+\frac{\rho}{2}\left\|\mathbf{X}_q-\mathbf{V}_q^{(n)}\right\|_F^2,
\label{eq:app_dc_subproblem}
\end{equation}
where $\mathbf{V}_q^{(n)}$ is defined in~\eqref{eq:sec3_v_time}. Setting the gradient to zero gives
\begin{equation}
\frac{1}{\sigma_q^2}\mathbf{F}_{\mathrm{sa}}^H\!\left(\mathbf{F}_{\mathrm{sa}}\mathbf{X}_q\mathbf{A}_q^H-\mathbf{Y}_q\right)\!\mathbf{A}_q+\rho\left(\mathbf{X}_q-\mathbf{V}_q^{(n)}\right)=\mathbf{0}.
\label{eq:app_gradient_zero}
\end{equation}
Using $\mathbf{F}_{\mathrm{sa}}^H\mathbf{F}_{\mathrm{sa}}=\mathbf{I}$ (unitarity of the spatial--angle DFT in the present $k_{\mathrm{sa}}=1$ setting) and rearranging:
\begin{equation}
\mathbf{X}_q\!\left(\frac{1}{\sigma_q^2}\mathbf{A}_q^H\mathbf{A}_q+\rho\,\mathbf{I}\right)=\frac{1}{\sigma_q^2}\mathbf{F}_{\mathrm{sa}}^H\mathbf{Y}_q\mathbf{A}_q+\rho\,\mathbf{V}_q^{(n)}.
\label{eq:app_rearranged}
\end{equation}
Let $\alpha\triangleq\rho\sigma_q^2$. Dividing both sides by $\rho$ yields
\begin{equation}
\mathbf{X}_q\!\left(\frac{1}{\alpha}\mathbf{A}_q^H\mathbf{A}_q+\mathbf{I}\right)=\frac{1}{\alpha}\mathbf{F}_{\mathrm{sa}}^H\mathbf{Y}_q\mathbf{A}_q+\mathbf{V}_q^{(n)}.
\label{eq:app_divided}
\end{equation}
Applying the Sherman--Morrison--Woodbury (SMW) identity to the right-hand inverse:
\begin{equation}
\left(\mathbf{I}+\frac{1}{\alpha}\mathbf{A}_q^H\mathbf{A}_q\right)^{-1}=\mathbf{I}-\frac{1}{\alpha}\mathbf{A}_q^H\!\left(\mathbf{I}+\frac{1}{\alpha}\mathbf{A}_q\mathbf{A}_q^H\right)^{-1}\!\mathbf{A}_q.
\label{eq:app_smw}
\end{equation}
Because $\mathbf{A}_q\in\mathbb{C}^{M_{\mathrm{p}}\times N_\tau}$ is a row selection of $\mathbf{F}_{\mathrm{fd}}$, its rows are orthonormal: $\mathbf{A}_q\mathbf{A}_q^H=\mathbf{I}_{M_{\mathrm{p}}}$. Substituting into~\eqref{eq:app_smw}:
\begin{equation}
\left(\mathbf{I}+\frac{1}{\alpha}\mathbf{A}_q^H\mathbf{A}_q\right)^{-1}=\mathbf{I}-\frac{\mathbf{A}_q^H\mathbf{A}_q}{\alpha+1}.
\label{eq:app_smw_simplified}
\end{equation}
Multiplying~\eqref{eq:app_divided} on the right by~\eqref{eq:app_smw_simplified} and writing $\mathbf{C}_q^{(n)}=\mathbf{V}_q^{(n)}+\frac{1}{\alpha}\mathbf{F}_{\mathrm{sa}}^H\mathbf{Y}_q\mathbf{A}_q$ gives the final closed-form update:
\begin{equation}
\mathbf{X}_q^{(n)}=\mathbf{C}_q^{(n)}-\frac{1}{\alpha+1}\,\mathbf{C}_q^{(n)}\mathbf{A}_q^H\mathbf{A}_q.
\label{eq:app_final}
\end{equation}
Eqs.~\eqref{eq:sec3_dc_closed_form}--\eqref{eq:sec3_dc_c} are recovered by substituting $\alpha=\rho\sigma_q^2$ back into $\mathbf{C}_q^{(n)}$ and the denominator. The key structural property exploited here is the row-orthonormality of $\mathbf{A}_q$ in~\eqref{eq:sec2_sensing_row_orth}, which reduces the $N_\tau\times N_\tau$ matrix inverse to a rank-$M_{\mathrm{p}}$ correction computable without any iterative solver.

\section{Training and Baseline Details}\label{app:training_details}

This appendix collects the training hyperparameters for DDA-Net and the configuration details of all baseline methods.

\subsection{DDA-Net Training Hyperparameters}
Table~\ref{tab:hyperparams} summarizes the full-data pretraining configuration for DDA-Net. All full-data variants (Doppler-3D, Time-3D, Time-2D) share the same optimizer and scheduling settings; the model-specific differences in kernel shape and temporal parameterization are described in Section~\ref{sec:impl_network} and Section~\ref{sec:ablation_model}. These models are trained until convergence via early stopping on the validation NMSE.

\begin{table}[t]
\centering
\caption{DDA-Net full-data pretraining hyperparameters.}
\label{tab:hyperparams}
\begin{tabular}{l l}
\toprule
Parameter & Value \\
\midrule
Optimizer & Adam ($\beta_1{=}0.9,\beta_2{=}0.999$) \\
Learning rate & $4\times 10^{-4}$ \\
Weight decay & $10^{-5}$ \\
Batch size & 8 \\
LR scheduler & ReduceLROnPlateau (factor 0.5, patience 6) \\
Early stopping & patience 15, min-delta $10^{-6}$ \\
Gradient clipping & max-norm 1.0 \\
GPU & NVIDIA Tesla V100 \\
\bottomrule
\end{tabular}
\end{table}

\subsection{Learning-Based Baseline Configurations}
\textbf{ChannelNet} uses a two-stage 3D CNN consisting of SRCNN3D followed by DNCNN3D, operating on the full time-frequency-space window with real/imaginary channel splitting and a pilot-mask input.

\textbf{LDAMP} uses a 2D slice-wise LDAMP formulation with a DnCNN denoiser, $10$ unrolled iterations, $32$ hidden channels, and $9$ convolutional layers per denoiser. It is applied independently to each time slice as described in Section~\ref{sec:impl_baselines}.

\subsection{Classical Baseline Configurations}

\begin{table}[t]
\centering
\caption{Classical baseline hyperparameters used in the evaluations.}
\label{tab:classical_params}
\begin{tabular}{l l}
\toprule
Method & Key parameters \\
\midrule
FISTA & $\lambda=0.1$--$1.2$, Lipschitz step, 20--200 iter \\
ADMM & $\rho=0.0125$--$1.0$, $\lambda=0.1$--$0.8$, 40--200 iter \\
Off-Grid-MS HMP & 80--640 iter, $\rho_{\mathrm{init}}=0.2$--$0.8$, \\
 & off-grid refinement and shared-MMV statistics \\
\bottomrule
\end{tabular}
\end{table}

\begin{table}[!t]
\centering
\caption{Fixed-speed UMa-NLOS NMSE (dB) at $10$~dB SNR for the 40-ms, $K=17$ setting, using $200$ samples per speed and all $17$ offsets. Bold/underline: best/best competing.}
\label{tab:fixed_speed_uma}
\footnotesize
\setlength{\tabcolsep}{0.8pt}
\renewcommand{\arraystretch}{0.95}
\textit{(a) Standard hopping pilot}\\[2pt]
\begin{tabular}{lrrrrrr}
\toprule
Method & \multicolumn{6}{c}{UE speed (km/h)} \\
\cmidrule(lr){2-7}
 & $0.1$ & $0.3$ & $0.5$ & $0.8$ & $1$ & $3$ \\
\midrule
LS & $-0.24$ & $-0.24$ & $-0.24$ & $-0.24$ & $-0.24$ & $-0.24$ \\
FISTA & $-1.16$ & $-0.45$ & $-0.13$ & $-0.63$ & $-0.61$ & $-0.42$ \\
ADMM & $-0.98$ & $-0.45$ & $-0.18$ & $-0.58$ & $-0.57$ & $-0.41$ \\
Off-Grid-MS HMP & $\underline{-2.07}$ & $\underline{-2.08}$ & $\underline{-2.06}$ & $\underline{-2.04}$ & $\underline{-2.02}$ & $\underline{-1.90}$ \\
ChannelNet & $-1.31$ & $-1.28$ & $-1.23$ & $-1.16$ & $-1.11$ & $-0.48$ \\
LDAMP & $-1.21$ & $-1.21$ & $-1.21$ & $-1.21$ & $-1.21$ & $-1.23$ \\
\textbf{DDA-Net} & $\mathbf{-9.06}$ & $\mathbf{-8.75}$ & $\mathbf{-8.46}$ & $\mathbf{-8.25}$ & $\mathbf{-8.06}$ & $\mathbf{-6.83}$ \\
\bottomrule
\end{tabular}

\vspace{4pt}
\textit{(b) MCD pilot}\\[2pt]
\begin{tabular}{lrrrrrr}
\toprule
Method & \multicolumn{6}{c}{UE speed (km/h)} \\
\cmidrule(lr){2-7}
 & $0.1$ & $0.3$ & $0.5$ & $0.8$ & $1$ & $3$ \\
\midrule
LS & $-0.24$ & $-0.24$ & $-0.24$ & $-0.24$ & $-0.24$ & $-0.24$ \\
FISTA & $-6.50$ & $-2.30$ & $-1.11$ & $-2.94$ & $-2.87$ & $-1.88$ \\
ADMM & $\underline{-6.85}$ & $-2.35$ & $-1.12$ & $-3.02$ & $-2.95$ & $-1.91$ \\
Off-Grid-MS HMP & $-5.69$ & $\underline{-5.72}$ & $\underline{-5.60}$ & $\underline{-5.52}$ & $\underline{-5.40}$ & $\underline{-4.72}$ \\
ChannelNet & $-0.46$ & $-0.48$ & $-0.49$ & $-0.54$ & $-0.56$ & $-0.33$ \\
LDAMP & $-1.21$ & $-1.21$ & $-1.21$ & $-1.21$ & $-1.21$ & $-1.23$ \\
\textbf{DDA-Net} & $\mathbf{-12.79}$ & $\mathbf{-12.51}$ & $\mathbf{-12.32}$ & $\mathbf{-12.08}$ & $\mathbf{-11.83}$ & $\mathbf{-10.38}$ \\
\bottomrule
\end{tabular}
\end{table}

Table~\ref{tab:classical_params} summarizes the classical baseline configurations used across domains, pilot settings, and representative SNR groups. Off-Grid-MS HMP retains only the estimation stage of Zhu~\emph{et al.} under our observation operator. It uses delay oversampling factors $2$--$3$, 80--640 message-passing iterations, off-grid refinement, and shared multi-measurement-vector statistics. Its main UMa-NLOS costs are reported in Table~\ref{tab:complexity}.

\section{Fixed-Speed UE Mobility Evaluation}\label{app:speed_sensitivity}

Because the main UMa-NLOS speed range ($1$--$4$~km/h) straddles the approximately $3.9$~km/h unaliased radial-speed threshold at $\Delta t=40$~ms, we additionally evaluate six fixed-speed test sets in the lower-speed, unaliased regime ($0.1$, $0.3$, $0.5$, $0.8$, $1$, and $3$~km/h; $200$ samples each). These test sets retain the main 40-ms, $K=17$ geometry, and the first four speeds are below the training range. The main-setting estimators are evaluated without retraining at $10$~dB under both pilots and all $17$ offsets.

Table~\ref{tab:fixed_speed_uma} reports the complete results. DDA-Net ranks first in all 12 speed--pilot combinations and improves monotonically at lower speeds: from $3$ to $0.1$~km/h, its NMSE improves by $2.23$~dB under Standard hopping and $2.41$~dB under MCD. Its margins over the strongest competitor remain $4.94$--$6.99$~dB under Standard hopping and $5.66$--$6.78$~dB under MCD. \mbox{ST-DDA} analysis shows that sufficiently small path-coefficient variation rates and phase curvature induce only limited additional Doppler spreading within a finite window~\cite{Zhu2026STDDA}; the observed trend is therefore consistent with greater DDA concentration at lower speeds despite within-window path evolution.

\bibliographystyle{IEEEtran}
\bibliography{IEEEabrv,references}

\begin{thebibliography}{10}
\providecommand{\url}[1]{#1}
\csname url@samestyle\endcsname
\providecommand{\newblock}{\relax}
\providecommand{\bibinfo}[2]{#2}
\providecommand{\BIBentrySTDinterwordspacing}{\spaceskip=0pt\relax}
\providecommand{\BIBentryALTinterwordstretchfactor}{4}
\providecommand{\BIBentryALTinterwordspacing}{\spaceskip=\fontdimen2\font plus
\BIBentryALTinterwordstretchfactor\fontdimen3\font minus
  \fontdimen4\font\relax}
\providecommand{\BIBforeignlanguage}[2]{{%
\expandafter\ifx\csname l@#1\endcsname\relax
\typeout{** WARNING: IEEEtran.bst: No hyphenation pattern has been}%
\typeout{** loaded for the language `#1'. Using the pattern for}%
\typeout{** the default language instead.}%
\else
\language=\csname l@#1\endcsname
\fi
#2}}
\providecommand{\BIBdecl}{\relax}
\BIBdecl

\bibitem{Marzetta2010MassiveMIMO}
T.~L. Marzetta, ``Noncooperative cellular wireless with unlimited numbers of
  base station antennas,'' \emph{IEEE Transactions on Wireless Communications},
  vol.~9, no.~11, pp. 3590--3600, 2010.

\bibitem{Larsson2014MassiveMIMO}
E.~G. Larsson, O.~Edfors, F.~Tufvesson, and T.~L. Marzetta, ``Massive {MIMO}
  for next generation wireless systems,'' \emph{IEEE Communications Magazine},
  vol.~52, no.~2, pp. 186--195, 2014.

\bibitem{Truong2013ChannelAging}
K.~T. Truong and J.~Robert W.~Heath, ``Effects of channel aging in massive
  {MIMO} systems,'' \emph{Journal of Communications and Networks}, vol.~15,
  no.~4, pp. 338--351, 2013.

\bibitem{GPP38331}
{3rd Generation Partnership Project (3GPP)}, ``{NR}; radio resource control
  (rrc); protocol specification,'' ETSI, Technical Specification TS 38.331,
  Version 18.6.0, Release 18, 2025.

\bibitem{GPP38211}
------, ``{NR}; physical channels and modulation,'' ETSI, Technical
  Specification TS 38.211, Version 18.5.0, Release 18, 2025.

\bibitem{Gao2016SCSMassiveMIMO}
Z.~Gao, L.~Dai, W.~Dai, B.~Shim, and Z.~Wang, ``Structured compressive
  sensing-based spatio-temporal joint channel estimation for {FDD} massive
  {MIMO},'' \emph{IEEE Transactions on Communications}, vol.~64, no.~2, pp.
  601--617, 2016.

\bibitem{Liu2016BurstSparsity}
A.~Liu, V.~K.~N. Lau, and W.~Dai, ``Exploiting burst-sparsity in massive {MIMO}
  with partial channel support information,'' \emph{IEEE Transactions on
  Wireless Communications}, vol.~15, no.~11, pp. 7820--7830, 2016.

\bibitem{Shen2019OTFSMassiveMIMO}
W.~Shen, L.~Dai, J.~An, P.~Fan, and J.~Robert W.~Heath, ``Channel estimation
  for orthogonal time frequency space ({OTFS}) massive {MIMO},'' \emph{IEEE
  Transactions on Signal Processing}, vol.~67, no.~16, pp. 4204--4217, 2019.

\bibitem{Yin2020PADPrediction}
H.~Yin, H.~Wang, Y.~Liu, and D.~Gesbert, ``Addressing the curse of mobility in
  massive {MIMO} with {Prony}-based angular-delay domain channel predictions,''
  \emph{IEEE Journal on Selected Areas in Communications}, vol.~38, no.~12, pp.
  2903--2917, 2020.

\bibitem{Chi2011BasisMismatch}
Y.~Chi, L.~L. Scharf, A.~Pezeshki, and A.~R. Calderbank, ``Sensitivity to basis
  mismatch in compressed sensing,'' \emph{IEEE Transactions on Signal
  Processing}, vol.~59, no.~5, pp. 2182--2195, 2011.

\bibitem{Wei2022OffGridOTFS}
Z.~Wei, W.~Yuan, S.~Li, J.~Yuan, and D.~W.~K. Ng, ``Off-grid channel estimation
  with sparse {B}ayesian learning for {OTFS} systems,'' \emph{IEEE Transactions
  on Wireless Communications}, vol.~21, no.~9, pp. 7407--7426, 2022.

\bibitem{Zhu2025FHSJCEP}
Y.~Zhu, J.~Zhuang, G.~Sun, H.~Hou, L.~You, and W.~Wang, ``Joint channel
  estimation and prediction for massive {MIMO} with frequency hopping
  sounding,'' \emph{IEEE Transactions on Communications}, vol.~73, no.~7, pp.
  5139--5154, 2025.

\bibitem{Tang2013OffGrid}
G.~Tang, B.~N. Bhaskar, P.~Shah, and B.~Recht, ``Compressed sensing off the
  grid,'' \emph{IEEE Transactions on Information Theory}, vol.~59, no.~11, pp.
  7465--7490, 2013.

\bibitem{Yang2016MDToeplitz}
Z.~Yang, L.~Xie, and P.~Stoica, ``Vandermonde decomposition of multilevel
  toeplitz matrices with application to multidimensional super-resolution,''
  \emph{IEEE Transactions on Information Theory}, vol.~62, no.~6, pp.
  3685--3701, 2016.

\bibitem{Wan2024TDD5GNRExtrapolation}
Y.~Wan and A.~Liu, ``A two-stage 2d channel extrapolation scheme for {TDD} 5g
  {NR} systems,'' \emph{IEEE Transactions on Wireless Communications}, vol.~23,
  no.~8, pp. 8497--8511, 2024.

\bibitem{Wan2025PilotPatternOptimization}
Y.~Wan, A.~Liu, and T.~Q.~S. Quek, ``Multi-user pilot pattern optimization for
  channel extrapolation in 5g {NR} systems,'' \emph{IEEE Transactions on
  Wireless Communications}, vol.~24, no.~7, pp. 6166--6179, 2025.

\bibitem{Ye2018PowerDL}
H.~Ye, G.~Y. Li, and B.-H.~F. Juang, ``Power of deep learning for channel
  estimation and signal detection in {OFDM} systems,'' \emph{IEEE Wireless
  Communications Letters}, vol.~7, no.~1, pp. 114--117, 2018.

\bibitem{Soltani2018ChannelNet}
M.~Soltani, V.~Pourahmadi, A.~Mirzaei, and H.~Sheikhzadeh, ``Deep
  learning-based channel estimation,'' \emph{IEEE Communications Letters},
  vol.~23, no.~4, pp. 652--655, 2019.

\bibitem{Luan2023Channelformer}
D.~Luan and J.~Thompson, ``Channelformer: Attention based neural solution for
  wireless channel estimation and effective online training,'' \emph{IEEE
  Transactions on Wireless Communications}, vol.~22, no.~10, pp. 6562--6577,
  2023.

\bibitem{Liu2024LBPCE}
C.~Liu, W.~Jiang, and X.~Yuan, ``Learning-based block-wise planar channel
  estimation for time-varying {MIMO} {OFDM},'' \emph{IEEE Wireless
  Communications Letters}, vol.~13, no.~8, pp. 2125--2129, 2024.

\bibitem{Chun2019MassiveMIMODLCE}
C.~J. Chun, J.~M. Kang, and I.~M. Kim, ``Deep learning-based channel estimation
  for massive {MIMO} systems,'' \emph{IEEE Wireless Communications Letters},
  vol.~8, no.~4, pp. 1228--1231, 2019.

\bibitem{He2019ModelDriven}
H.~He, S.~Jin, C.-K. Wen, F.~Gao, G.~Y. Li, and Z.~Xu, ``Model-driven deep
  learning for physical layer communications,'' \emph{IEEE Wireless
  Communications}, vol.~26, no.~5, pp. 77--83, 2019.

\bibitem{Balatsoukas2019DeepUnfolding}
A.~Balatsoukas-Stimming and C.~Studer, ``Deep unfolding for communications
  systems: A survey and some new directions,'' in \emph{2019 IEEE International
  Workshop on Signal Processing Systems ({SiPS})}, 2019, pp. 266--271.

\bibitem{Borgerding2017LAMP}
M.~Borgerding, P.~Schniter, and S.~Rangan, ``{AMP}-inspired deep networks for
  sparse linear inverse problems,'' \emph{IEEE Transactions on Signal
  Processing}, vol.~65, no.~16, pp. 4293--4308, 2017.

\bibitem{Metzler2017LDAMP}
C.~A. Metzler, A.~Mousavi, and R.~G. Baraniuk, ``Learned {D}-{AMP}: Principled
  neural network based compressive image recovery,'' in \emph{Advances in
  Neural Information Processing Systems 30}, 2017.

\bibitem{Yang2020ADMMCSNet}
Y.~Yang, J.~Sun, H.~Li, and Z.~Xu, ``{ADMM-CSNet}: A deep learning approach for
  image compressive sensing,'' \emph{IEEE Transactions on Pattern Analysis and
  Machine Intelligence}, vol.~42, no.~3, pp. 521--538, 2020.

\bibitem{He2018LDAMP}
H.~He, C.-K. Wen, S.~Jin, and G.~Y. Li, ``Deep learning-based channel
  estimation for beamspace mmwave massive {MIMO} systems,'' \emph{IEEE Wireless
  Communications Letters}, vol.~7, no.~5, pp. 852--855, 2018.

\bibitem{Boyd2011ADMM}
S.~Boyd, N.~Parikh, E.~Chu, B.~Peleato, and J.~Eckstein, ``Distributed
  optimization and statistical learning via the alternating direction method of
  multipliers,'' \emph{Foundations and Trends in Machine Learning}, vol.~3,
  no.~1, pp. 1--122, 2011.

\bibitem{Zhu2026STDDA}
\BIBentryALTinterwordspacing
X.~Zhu and T.~Li, ``{ST-DDA}: Dynamic channel estimation in the
  doppler-delay-angle domain via sparse subspace tracking for {TDD} systems,''
  2026. [Online]. Available: \url{https://arxiv.org/abs/2607.19677}
\BIBentrySTDinterwordspacing

\bibitem{Jaeckel2014QuaDRiGa}
S.~Jaeckel, L.~Raschkowski, K.~B{\"o}rner, and L.~Thiele, ``{QuaDRiGa}: A 3-{D}
  multi-cell channel model with time evolution for enabling virtual field
  trials,'' \emph{IEEE Transactions on Antennas and Propagation}, vol.~62,
  no.~6, pp. 3242--3256, 2014.

\bibitem{GPP38901}
{3rd Generation Partnership Project (3GPP)}, ``Study on channel model for
  frequencies from 0.5 to 100 {GHz},'' ETSI, Technical Report TR 38.901,
  Version 18.0.0, Release 18, 2024.

\bibitem{Zhu2026MCCPilot}
X.~Zhu, Y.~Zeng, and T.~Li, ``Coverage-distance and collinearity-minimizing
  pilots for channel estimation in {TDD} systems,'' \emph{IEEE Wireless
  Communications Letters}, vol.~15, pp. 3741--3745, 2026.

\bibitem{Beck2009FISTA}
A.~Beck and M.~Teboulle, ``A fast iterative shrinkage-thresholding algorithm
  for linear inverse problems,'' \emph{SIAM Journal on Imaging Sciences},
  vol.~2, no.~1, pp. 183--202, 2009.

\end{thebibliography}

\end{document}